\documentclass[10pt]{article}
\usepackage{amssymb}
\usepackage{amsmath}
\usepackage{achemso}
\usepackage{graphicx}

\input{tcilatex}

\begin{document}

{\huge \ }

{\huge New Strings for Old Veneziano }

{\huge Amplitudes \ III. Symplectic Treatment}

$\ \ \ \ \ \ \ \ \ \ \ \ \ \ \ \ \ \ \ \ \ \ \ \ \ \ \ \ \ \ \ \ \ $

\ \ \ \ \ \ \ \ \ \ \ \ \ \ \ \ \ \ \ \ \ \ \ \ \ \ \ A.L. Kholodenko

\textit{375 H.L.Hunter Laboratories, Clemson University, Clemson, }

\textit{SC} 29634-0973, USA

\bigskip\textbf{Abstract}

A $d-$dimensional rational polytope $\mathcal{P}$ is a polytope whose
vertices are located at the nodes of \textbf{Z}$^{d}$ lattice. Consider the
number $\left\vert k\mathcal{P}\cap \mathbf{Z}^{d}\right\vert $ of points
inside the inflated $\mathcal{P}$ with coefficient of inflation $k$ $%
(k=1,2,3,...).$ The Ehrhart polynomial of $\mathcal{P}$ counts the number of
such lattice points inside the inflated $\mathcal{P}$ and (may be) at its
faces (including vertices). In Part I [ JGP 55 (2005) 50] of our four parts
work we noticed that Veneziano amplitude is just the Laplace transform of
the generating function (considered as a partition function in the sence of
statistical mechanics) for the Ehrhart polynomial for the regular inflated
simplex obtained as deformation retract of the Fermat (hyper) surface living
in the complex projective space. This observation is sufficient for
development of \ new symplectic (this work) and supersymmetric (Part II)
physical models reproducing the Veneziano (and Veneziano-like) amplitudes.
General ideas (e.g. those related to the properties of Ehrhart polynomials)
are illustrated by simple practical examples (e.g. use of mirror symmetry
for explanation of available experimental data on $\pi \pi $ scattering ,
etc.) worked out in some detail. Obtained final results are in formal accord
with those earlier obtained by Vergne [PNAS 93 (1996)14238 ].

\bigskip


\ 

\textit{Subj Class}.: String theory; Torus actions on symplectic manifolds;
Combinatorics of polytopes; Weyl character formula

\ 


\bigskip

\pagebreak

\ \ \ 

\section{\protect\bigskip Introduction}

\subsection{Connection with earlier work}

\bigskip

In \ our earlier works, Refs.[1,2], which we shall call Part I and II%
\footnote{%
In referring to the results of these papers we shall use notations like
Eq.(II. 5.10), etc.}, we initiated development of new formalism reproducing
both the Veneziano and Veneziano-like (tachyon-free) amplitudes and models
generating these amplitudes. \ In particular, in Part II\ we discussed one
of such models. Contrary to traditional treatments, we demonstrated that our
model is supersymmetric and finite-dimensional. This result was obtained
with help of the theory of invariants of pseudo-reflection groups. The
partition function, Eq.(II,6.10), for this model is given by the Poincare$%
^{\prime }$ polynomial%
\begin{equation}
P((S(V)\otimes E(V))^{G};z)=\prod\limits_{i=1}^{n}\frac{1-z^{q+i}}{1-z^{i}}.
\tag{1.1}
\end{equation}%
In the limit: $z\rightarrow 1,$ the above result is reduced to 
\begin{equation}
P((S(V)\otimes E(V))^{G};\text{ }z=1)=\frac{(q+1)(q+2)\cdot \cdot \cdot (q+n)%
}{n!}\equiv p(q,n).  \tag{1.2}
\end{equation}%
which is Eq.(II,6.11). The detailed combinatorial explanation of these
results was given already in Part II. In this work, to avoid repetitions, we
would like to extend \ such an explanation having in mind development of the
symplectic model reproducing Veneziano amplitudes.

Steps toward designing of such a model were made already in Part I where it
was noticed that the unsymmetrized Veneziano amplitude is obtainable as the
Laplace transform of the partition function 
\begin{equation}
P(q,t)=\sum\limits_{n=0}^{\infty }p(q,n)t^{n}  \tag{1.3}
\end{equation}%
where\ $p(q,n)$ is the same as in Eq.(1.2). In Ref.[3] Vergne demonstrated
(without reference to string theory or Veneziano amplitudes) that such
partition function has both symplectic and quantum mechanical meaning : the
quantity $p(q,n)$ is dimension of the quantum Hilbert space associated
(through the coadjoint orbit method) with the classical system made out of \ 
\textit{finite} number of harmonic oscillators living on a specially
designed symplectic manifold.

In this work using different arguments we reobtain her final results. Our
use of different arguments is motivated by our desire to demonstrate
connections between the formalism developed in this paper and that already
in use in the mathematical physics literature. More importantly, the
treatment presented below complements that developed earlier in Parts I and
II.

In Part II, following work by Lerche et al [4], we adopted the idea that 
\textit{any} kind of one variable Poincare$^{\prime }$ polynomial (actually,
up to a constant) can be interpreted as the Weyl character formula. Since,
according to Part II, both Eq.s (1.1) and (1.3) are Poincare$^{\prime }$
polynomials, their interpretation in terms of the Weil character formula
provides major ingredient toward reconstruction of the Veneziano amplitudes
from \ the underlying quantum mechanical partition function (the Weyl
character formula). Going into opposite direction, such amplitudes acquire
some topological meaning to be further illuminated in this work. Direct link
between topology (the Poincare$^{\prime }$ polynomials) and quantum
mechanics (the Weyl character formula) is certainly not limited to its use
only for the Veneziano amplitudes and is of independent interest. In view of
this, in the next subsection we would like to provide simple arguments
(different from those in the work by Lerche et al) explaining why this is so.

\subsection{A motivating example}

\bigskip Consider a finite geometric progression of the type 
\begin{align}
\mathcal{F(}c,m)& =\sum\limits_{l=-m}^{m}\exp \{cl\}=\exp
\{-cm\}\sum\limits_{l=0}^{\infty }\exp \{cl\}+\exp
\{cm\}\sum\limits_{l=-\infty }^{0}\exp \{cl\}  \notag \\
& =\exp \{-cm\}\frac{1}{1-\exp \{c\}}+\exp \{cm\}\frac{1}{1-\exp \{-c\}} 
\notag \\
& =\exp \{-cm\}\left[ \frac{\exp \{c(2m+1)\}-1}{\exp \{c\}-1}\right] . 
\tag{1.4}
\end{align}%
The reason for displaying the intermediate steps will become apparent
shortly. First, however, we would like to consider the limit : $c\rightarrow
0^{+}$ of $\mathcal{F(}c,m)$. Clearly, it is given by $\mathcal{F(}0,m)=2m+1$%
. The number $2m+1$ equals to the number of integer points in the segment $%
[-m,m]$ \textit{including} \textit{boundary} points. It is convenient to
rewrite the above result in terms of $x=\exp \{c\}$. We shall write formally 
$\mathcal{F(}x,m)$ instead of $\mathcal{F(}c,m)$ from now on. Using these
notations, let us consider the related function, 
\begin{equation}
\mathcal{\bar{F}(}x,m)=(-1)\mathcal{F(}\frac{1}{x},-m).  \tag{1.5}
\end{equation}%
Such type of relation (the \textit{Ehrhart-Macdonald reciprocity law}) is
characteristic for the Ehrhart polynomial for the rational polytopes to be
defined in the next subsection. In Ref.[5] Stanley provides many
applications of this reciprocity law. In our case, we obtain explicitly, 
\begin{equation}
\mathcal{\bar{F}(}x,m)=(-1)\frac{x^{-(-2m+1)}-1}{x^{-1}-1}x^{m}.  \tag{1.6}
\end{equation}%
In the limit $x\rightarrow 1+0^{+}$ we obtain : $\mathcal{\bar{F}(}1,m)=2m-1.
$ The number $2m-1$ is equal to the number of integer points strictly 
\textit{inside} the segment $[-m,m].$ These, seemingly trivial, results can
be broadly generalized. First, we replace $x$ by \textbf{x.} Next, we
replace the summation sign in the left hand side of Eq.(1.4) by the multiple
summation, etc. Thus obtained function $\mathcal{F(}\mathbf{x},m)$ in the
limit $x_{i}\rightarrow 1+0^{+},$ $i=1-d,$ produces the anticipated result: 
\begin{equation}
\mathcal{F(}\mathbf{1},m)=(2m+1)^{d}  \tag{1.7}
\end{equation}%
for number of points inside and at the edges of the $d$ dimensional cube in
Euclidean space \textbf{R}$^{d}$. Accordingly, for the number of points
strictly inside the cube we obtain: $\mathcal{\bar{F}(}1,m)=(2m-1)^{d}.$ The
rationale for describing this limiting procedure is caused by its connection
with our earlier result, Eq.(1.2). To explain this we need to extend our
simple results in order to describe analogous situation for arbitrary
centrally symmetric polytope. This is accomplished in several steps. We
begin with some definitions.

\textbf{Definition 1.1.} A subset of \ \textbf{R}$^{n}$ is a \textit{%
polytope (or polyhedron)} $\mathcal{P}$ if there is a $r\times d$ matrix $%
\mathbf{M}$\textbf{\ (}with\textbf{\ }$r\leq d)$ and a vector $\mathbf{b}\in 
\mathbf{R}^{d}$ such that 
\begin{equation}
\mathcal{P}=\{\mathbf{x}\in\mathbf{R}^{d}\mid\mathbf{Mx}\leq\mathbf{b}\}. 
\tag{1.8}
\end{equation}

\textbf{Definition 1.2. }Provided that the Euclidean $d$-dimensional scalar
product is given by 
\begin{equation}
<\mathbf{x}\cdot \mathbf{y}>=\sum\limits_{i=1}^{d}x_{i}y_{i}  \tag{1.9}
\end{equation}%
\footnote{%
So that \textbf{x }lives in space \textit{dual} to that for \textbf{y}.}a 
\textit{rational (}respectively\textit{, integral) }polytope\textit{\ (or
polyhedron)} $\mathcal{P}$ is defined by the set 
\begin{equation}
\mathcal{P}=\{\mathbf{x}\in \mathbf{R}^{d}\mid <\mathbf{a}_{i}\cdot \mathbf{x%
}>\leq \beta _{i}\text{ , }i=1,...,r\}  \tag{1.10}
\end{equation}%
where $\ \mathbf{a}_{i}\in \mathbf{Q}^{n}and$ $\beta _{i}\in \mathbf{Q}$ for 
$i=1,...,r\ ($respectively, $\mathbf{a}_{i}\in \mathbf{Z}^{n}$ and $\beta
_{i}\in \mathbf{Z}$ for $i=1,...,r).$

Let \textit{Vert}$\mathcal{P}$ denote the vertex set of the rational
polytope, in the case considered thus far, the $d-$dimensional cube. Let $%
\{u_{1}^{v},...,u_{d}^{v}\}$ be the orthogonal basis (not necessarily of
unit length) made of the highest weight vectors of the Weyl-Coxeter
reflection group $B_{d}$ appropriate for the cubic symmetry\footnote{%
For a brief guide to the Weyl-Coxeter reflection groups, please, see
Appendix to Part II}.These vectors are oriented along the positive semi axes
with respect to the center of symmetry of (hyper)cube. When parallel
translated to the edges ending at particular hypercube vertex \textbf{v},
they can point either in or out of this vertex. In terms of these notations,
the $d$-dimensional version of Eq.(1.4) can be rewritten now as follows 
\begin{equation}
\sum\limits_{\mathbf{x}\in \mathcal{P\cap }\mathbf{Z}^{d}}\exp \{<\mathbf{c}%
\cdot \mathbf{x}>\}=\sum\limits_{\mathbf{v}\in Vert\mathcal{P}}\exp \{<%
\mathbf{c}\cdot \mathbf{v}>\}\left[ \prod\limits_{i=1}^{d}(1-\exp
\{-c_{i}u_{i}^{v}\})\right] ^{-1}.  \tag{1.11}
\end{equation}%
Correctness of this equation can be readily checked by considering special
cases of a segment, square, cube, etc. The result, Eq.(1.11), obtained for
the polytope of cubic symmetry can be extended to the arbitrary convex
centrally symmetric polytope as we shall demonstrate below. This fact allows
us to investigate properties of more complex polytopes with help of
polytopes of cubic symmetry. Moreover, we shall argue below that the r.h.s.
of Eq.(1.11) is mathematically equivalent to the r.h.s. of Eq.(1.1). Because
of this, the limiting procedure $c\rightarrow 0^{+}$ producing the number of
points inside (and at the boundaries) of the polyhedron $\mathcal{P}$ in the
l.h.s. of Eq.(1.11) is of the same nature as the limiting procedure:$%
z\rightarrow 1$ in Eq.(1.2) where, as result of this procedure, the r.h.s of
Eq.(1.2) produces the number of lattice points for the inflated (with
inflation coefficient $q$) rational simplex of dimension $n$
\textquotedblright living\textquotedblright\ in \textbf{Z}$^{n}$ lattice.

\textbf{Remark 1.3.} For an arbitrary convex polytope the above formula,
Eq.(1.11), was obtained (seemingly independently) in many different
contexts. For instance, in the context of discrete and computational
geometry it is attributed to Brion [6]. In view of \ Eq.(1.5), it can as
well be attributed to Ehrhart, Stanley [5] \ and to many others. In fact,
for the case of centrally symmetric polytopes this formula is just a special
case of the Weyl's character formula. This will be demonstrated below, in
Section 2.

There are many paths to arrive at final results of this paper, i.e. to
reobtain the results of Vergne [3], and to use them for construction of new
models reproducing the Veneziano and Veneziano-like amplitudes. They
include, for instance, the algebro-geometric, symplectic, group-theoretic,
combinatorial, supersymmetric, etc. pathways to reach the same destination.
In our opinion, the most direct way to arrive at final results is
combinatorial. Although it will be discussed at length in in Part IV from
yet another perspective, in this work we present some essentials needed for
their immediate use in the rest of the paper. In particular, we would like
to discuss now some facts about the Ehrhart polynomials having in mind their
uses in high energy physics.

\subsection{ Ehrhart polynomials, mirror symmetry and the extended Veneziano
amplitudes}

In the previous subsection we introduced Eq.(1.5). It characterizes the
Ehrhart polynomial. It is important to realize that $p(q,n)$ in Eq.(1.2)
already is an example of the Ehrhart polynomial. Evidently, Eq.(1.2) can be
written formally as 
\begin{equation}
p(q,n)=a_{n}q^{n}+a_{n-1}q^{n-1}+\cdot \cdot \cdot +a_{0}.  \tag{1.12}
\end{equation}%
In Ref.[7] it is argued that for \textit{any} convex rational polytope $%
\mathcal{P}$ the Ehrhart polynomial can be written as 
\begin{equation}
\left\vert q\mathcal{P}\cap \mathbf{Z}^{n}\right\vert =\mathfrak{P}%
(q,n)=a_{n}(\mathcal{P})q^{n}+a_{n-1}(\mathcal{P})q^{n-1}+\cdot \cdot \cdot
+a_{0}(\mathcal{P})  \tag{1.13}
\end{equation}%
with coefficients $a_{0},...,a_{n}$ specific for a given polytope $\mathcal{P%
}$. Nevertheless, irrespective to the type of polytope $\mathcal{P}$, it is
known that $a_{0}=1$ and $a_{n}=Vol\mathcal{P},$where $Vol\mathcal{P}$ is
the $\mathit{Euclidean}$ volume of the polytope. To calculate the remaining
coefficients of this polynomial explicitly for an arbitrary convex polytope
is a difficult task in general. Such task was accomplished rather recently
in Ref.[8].The authors of [8] recognized that in order to obtain the
remaining coefficients it is useful to calculate the generating function for
the Ehrhart polynomial. In our case this function is given by Eq.(1.3). From
Eq.(I,1.22) we already know that formally this is the partition function for
the unsymmetrized Veneziano amplitude. In view of our Eq.(1.1) taken from
Part II, we also know that it can be also looked upon as the partition
function for theVeneziano amplitudes. Hence, now we would like to explain
how Eq.s (1.1) and (1.3) are related to each other from the point of view of
commutative algebra and combinatorics of polytopes.By doing so some useful
physical information will be obtained as well.

Long before the results of Ref.[8] were published, it was known [9] that the
generating function for the Ehrhart polynomial of $\mathcal{P}$ can be
written in the following universal form 
\begin{equation}
\mathcal{F}(\mathcal{P},x)=\sum\limits_{q=0}^{\infty }\mathfrak{P}(q,n)x^{q}=%
\frac{h_{0}(\mathcal{P})+h_{1}(\mathcal{P})x+\cdot \cdot \cdot +h_{n}(%
\mathcal{P})x^{n}}{(1-x)^{n+1}}.  \tag{1.14}
\end{equation}%
For the Veneziano partition function all coefficients, except $h_{0}(%
\mathcal{P})$, are zero and, of course, $h_{0}(\mathcal{P})=1$ $[10].$ This
can be easily understood in view of Eq.(I.1.20). We brought to our readers
attention the above general result in view of our task of comparing
Eq.s(1.1) and (1.3). and of possibly generalizing the Veneziano amplitudes
and the partition functions associated with them.

In practical applications it should be noted that the combinatorial factor $%
p(q,n),$Eq$.(1.2),$ representing the number of points in the inflated
simplex \ $\mathcal{P}$ (with coefficient of inflation $q$) \ whose vertex
set \textit{Vert}$\mathcal{P}$ belongs to \textbf{Z}$^{n}$ lattice can be
formally written in several equivalent ways. In particular, as we've
mentioned in Part II, 
\begin{equation}
p(q,n)=\frac{(q+n)!}{q!n!}=\frac{(q+1)\cdot \cdot \cdot (q+n)}{n!}=\frac{%
(n+1)\cdot \cdot \cdot (n+q)}{q!}.  \tag{1.15}
\end{equation}%
This fact has some physical significance. For instance, in particle physics
literature, e.g. see Refs.[11,12], the third option is commonly used. Let us
recall how this happens. One is looking for an expansion of the factor $%
(1-x)^{-\alpha (t)-1}$ under the integral of beta function as explained in
Part I. \ Looking at Eq.(1.14) one realizes that the Mandelstam variable $%
\alpha (t)$ plays a role of dimensionality of \textbf{Z-} lattice. Hence, we
have to identify it with $n$ in the \textit{second option} provided by
Eq.(1.15). This is not the way such an identification is done in physics
literature where, in fact, the third option provided by Eq.(1.15) is
commonly used with $n=\alpha (t)$ effectively being the inflation factor
while $q$ effectively being the dimensionality of the lattice.\footnote{%
We have to warn our readers that, to our knowledge, nowhere in physics
literature such combinatorial terminology is being used.} A quick look at
Eq.s(1.3) and (1.14) shows that under such circumstances the generating
function for the Ehrhart polynomial and that for the Veneziano amplitude are
formally \textit{not} the same: in the first (mathematical) case one is
dealing with lattices of fixed dimensionality and is considering summation
over various inflation factors at the same time, while in the third
(physical) case, one is dealing with the fixed inflation factor $n=\alpha (t)
$ while summing over lattices of different dimensionalities. Such arguments
are superficial however in view of \ Eq.s(I.1.20) and (1.14) above. Using
these equations it is clear that correct agreement between Eq.s(1.3) and
(1.14) can be reached if one is using $\mathfrak{P}(q,n)=p(q,n)$ with the
second (i.e.mathematical) option offered by Eq.(1.15). By doing so no
changes in the pole locations for the Veneziano amplitude occur. Moreover,
for a given pole the second and the third option in Eq.(1.15) produce
exactly the same contributions into the residue thus making them \textit{%
physically} indistinguishable. Nevertheless, our choice of the
mathematically meanigful interpretation of the Veneziano amplitude as the
Laplace transform of the Ehrhart polynomial generating function provides one
of the $\mathit{major}$ $\mathit{reasons}$ for development of the formalism
of Parts I-through IV. In particular, it allows us to think about possible
generalizations of the Veneziano amplitude using generating functions for
the Ehrhart polynomials for polytopes \ of other types. As it is
demonstrated by Stanley [9,13], both Eq.(1.1) and Eq.(1.14) have group-
invariant meaning as Poincare$^{\prime }$ polynomials: Eq.(1.14) is
associated with the Poincare$^{\prime }$ polynomial for the so called
Stanley-Reisner polynomial ring while Eq.(1.1) is the Poincare polynomial
for the so called Gorenstein ring. Naturally, these two rings are
interrelated thus providing the desired connection between Eq.s(1.1) and
(1.14). For the sake of space, we refer our readers to the original works by
Stanley [9,13] where all mathematical details can be found. At the same
time, we have supplied sufficient information in order to discuss some
physical applications. In particular, following Batyrev, Ref.[14, p.392 ],
and Hibi, Ref.[15], we would like to discuss the reflexive (polar(or dual))
polytopes playing major role in calculations involving mirror symmetry. To
those\ of our readers who are familiar with some basic facts of solid state
physics [16] the concept of a dual (or polar) polytope should look quite
familiar since it is completely analogous to that for the reciprocal
lattice. Both direct and reciprocal lattices are used rutinely in
calculations related to physical properties of crysalline solids. The
requirement that physical observables should remain the same irrespective to
what lattice is used in computations is completely natural. Not
surprisingly, such a requirement \ formally coincides with that used in the
high energy physics. In the paper by Greene and Plesser, Ref.[17, p.26], one
finds the following statement: "Thus, we have demonstrated that two
topologically distinct Calabi-Yau manifolds $M$ and $M^{\prime }$ give rise
to the same conformal field theory. Furthermore, although our argument has
been based only at one point in the respective moduli spaces $\mathcal{M}_{M}
$ and $\mathcal{M}_{M^{\prime }}$ of $M$ and $M^{\prime }$(namely the point
which has a minimal model interpretation and hence respects the symmetries
by which we have orbifolded) the result necessarily extends to all of $%
\mathcal{M}_{M}$ and $\mathcal{M}_{M^{\prime }}$ ". In parts I and II we
argued that, in view of the Veneziano condition, there is a significant
difference between calculations of observables (amplitudes) of high energy
physics and those in conformal field theories. This difference is analogous
to the difference between the point group symmetries in the liquid/gas
phases and translational symmetries of solid phases. Hence, extension of the
Veneziano amplitudes with help of general result, Eq.(1.14), (which is
essentially equivalent to accounting for the mirror symmetry) requires some
explanations. We would like to to provide a sketch of these explanations now%
\footnote{%
To keeep focus of our readers on major issues of this paper, we supress to
the absolute minimum the discussion connecting our results to experiment. We
plan to discuss this connection thoroughly in a separate publication.}.

To this purpose we need to introduce several definitions first. We begin with

\bigskip \textbf{Definition 1.4}. For any convex polytope $\mathcal{P}$ the
dual polytope $\mathcal{P}^{\ast }$ is defined by\ \ \ \ \ \ \ \ \ \ \ \ \ \
\ \ \ \ \ \ \ \ \ \ \ \ \ \ \ \ \ \ \ \ \ \ \ \ \ \ \ \ \ \ \ \ \ \ \ \ \ \
\ \ \ \ \ \ \ \ 

\begin{equation}
\mathcal{P}^{\ast }=\{\mathbf{x}\in \left( \mathbf{R}^{d}\right) ^{\ast
}\mid <\mathbf{a}\cdot \mathbf{x}>\leq 1,\mathbf{a}\in \mathcal{P}\} 
\tag{1.16}
\end{equation}%
Although in algebraic geometry of toric varieties the inequality $<\mathbf{a}%
\cdot \mathbf{x}>\leq 1$ sometimes is replaced by $<\mathbf{a}\cdot \mathbf{x%
}>\geq -1$ [10], we shall use Eq.(1.16) to be in accord with Hibi, Ref.[15].
According to this reference, if $\mathcal{P}$ is rational, then $\mathcal{P}%
^{\ast }$ is also rational. Hovever, $\mathcal{P}^{\ast }$ is not
necessarily integral even if $\mathcal{P}$ is integral.This fact is
profoundly important since the result, Eq.(1.14), is valid for the integral
polytopes only.The question arises : under what conditions is the dual
polytope $\mathcal{P}^{\ast }$ inegral? The aswer is given by the following

\textbf{Theorem 1.5}.(Hibi, Ref.[15]). \textit{The dual polytope} $\mathcal{P%
}^{\ast }$ \textit{is \textbf{integral} if and only if}

\bigskip 
\begin{equation}
\mathcal{F}(\mathcal{P},x^{-1})=(-1)^{d+1}x\mathcal{F}(\mathcal{P},x) 
\tag{1.17}
\end{equation}%
\textit{where the generating function }$\mathcal{F}(\mathcal{P},x)$ \textit{%
is defined earlier by} Eq.(1.14).

By combining Eq.s(1.3) and (1.14) we obtain for the standard Veneziano
amplitude the following result:%
\begin{equation}
\mathcal{F}(\mathcal{P},x)=\left( \frac{1}{1-x}\right) ^{d+1}.  \tag{1.18}
\end{equation}%
Using it in Eq.(1.17) produces:%
\begin{equation}
\mathcal{F}(\mathcal{P},x^{-1})=\frac{(-1)^{d+1}}{(1-x)^{d+1}}%
x^{d+1}=(-1)^{d+1}x^{d+1}\mathcal{F}(\mathcal{P},x).  \tag{1.19}
\end{equation}%
This result indicates that scattering processes described by the standard
Veneziano amplitudes \textit{do not} involve any mirror symmetry since, as
it is well known, Refs [14,18], in order for such a symmetry to take place
the dual polytope $\mathcal{P}^{\ast }$ \textit{must} be integral. The
question arises: Can these amplitudes be modified with help of Eq.(1.14) so
that mirror symmetry can be \ in principle observed in nature? To answer
this question, let us assume that, indeed, Eq.(1.14) can be used for such a
modification. In this case we obtain 
\begin{equation}
\mathcal{F}(\mathcal{P},x^{-1})=(-1)^{d+1}x\mathcal{F}(\mathcal{P},x) 
\tag{1.20}
\end{equation}%
porovided that $h_{n-i}$ $=h_{i}$ in Eq.(1.14). But this is surely the case
in view of the fact that these are the Dehn-Sommerville equations, Ref.[10,
p.16]. Hence, at this stage of our discussion, it looks like generalization
of the Veneziano amplitudes which takes into account mirror symmetry is
possible from the mathematical standpoint. Mathematical arguments themself
are \textit{not sufficient} however for such generalization. This is so
because of the following chain of arguments.

Already in the original paper by Veneziano, Ref.[19, p.195], it was noticed
that the amplitude he defined originally is not unique. Following Ref.[20,
p.100], we notice that beta function $B(-\alpha (s),-\alpha (t))$ in
Veneziano amplitude can be replaced by%
\begin{equation}
B(m-\alpha (s),n-\alpha (t))  \tag{1.21}
\end{equation}%
for any integers $m,n\geq 1$, and similarly for $s,u$ and $t,u$ terms.
According to Ref. [20],  "Any function which can be presented as linear
combination of terms like (1.21) satisfies the finite energy sum rules, so
there is no constraint on the resonance parameters \textit{unless additional
assumptions are made}." The mirror symmetry arguments presented above 
\textit{are such additional assumptions}. To use them wisely, we still need
to make several remarks. Experimentally, the linear combination of terms in
the form given by Eq.(1.21) should show up in the form of daughter (or
satellite) Regge trajectories as explained inRef.[20,21]. But, according to
Frampton, Ref.[22], such daughter trajectories should be present even for
the standard (that is non generalized!) Veneziano amplitudes. This is so
because of the following arguments. In accord with analysis made in Part I,
the unsymmetrized Veneziano amplitude can be presented as 
\begin{equation}
V(s,t)=\dsum\limits_{n=0}^{\infty }p(\alpha (t),n)\frac{1}{n-\alpha (s)}. 
\tag{1.22}
\end{equation}%
For a given $n$ the polynomial $p(\alpha (t),n)$ is an $n-th$ degree
polynomial in $\alpha (t)$ $($for high enough energies in $t$). Since in the
Regge theory $n$ represents the total spin, according to the rules of
quantum mechanics, in addition to particles with spin $n$ there should
(could) be particles with spins $n-1$, $n-2,...,1,0.$These particles should
be visible (in principle) at the parallel (daughter or satellite)
trajectories all lying below the leading (with spin $n$) Regge trajectory.
While the leading trajectory has $\alpha (t)^{n}$ as its residue the daugher
trajectories have $\left( \alpha (t)\right) ^{n-1},\left( \alpha (t)\right)
^{n-2},$etc.as their residues. Unfortunately, in addition, the countable
infinity of satellite (daughter) trajectories can originate if the masses of
colliding particles are not the same, Ref.[20, p. 40].The linear combination
of terms given by Eq.(1.21) can account in principle for such phenomena.
Following Frampton, Ref.[22], we need to take into account that the linear
combination of terms in Eq.(1.21) can be replaced (quite rigorously) by the
combination of terms of the form%
\begin{equation}
B(n-\alpha (s),n-\alpha (t))  \tag{1.23}
\end{equation}%
with $n\geq 0.$In real life the infinite number of trajectories is never
observed however. But several parent-daughter Regge trajectories are being
observed quite frequently, e.g. see Ref.[20, p.41]. If we accept the
existence of mirror symmetry these observational facts can be explained
quite naturally.\ To illustrate this, we would like to consider the simplest
case of $\pi \pi $ scattering desribed in Ref.s[12,22]. Below the threshold,
that is below the collision energies producing more outgoing particles than
incoming, the unsymmetrized amplitude $A(s,t)$ \ for such a process is known
to be%
\begin{eqnarray}
A(s,t) &=&-g^{2}\frac{\Gamma (1-\alpha (s))\Gamma (1-\alpha (t))}{\Gamma
(1-\alpha (s)-\alpha (t))}  \notag \\
&=&-g^{2}(1-\alpha (s)-\alpha (t))B(1-\alpha (s),1-\alpha (t)). 
\TCItag{1.24}
\end{eqnarray}%
This result should be understood as follows. Consider the "weighted" \
(unsymmetrized) Veneziano amplitude of the type%
\begin{equation}
A(s,t)=\dint\limits_{0}^{1}dxx^{-\alpha (s)-1}(1-x)^{-\alpha (t)-1}g(x,s,t) 
\tag{1.25}
\end{equation}%
where the weight function $g(x,s,t)$ is given by%
\begin{equation}
g(x,s,t)=\frac{1}{2}g^{2}[(1-x)\alpha (s)+x\alpha (t)].  \tag{1.26a}
\end{equation}%
Alternatively, the same result, Eq.(1.24), is obtained if one uses instead
the weight function 
\begin{equation}
g(x,s,t)=g^{2}x\alpha (t)  \tag{1.26b}
\end{equation}%
Consider now a special case of Eq.(1.14): $n=2$. For such case we obtain 
\begin{equation}
\mathcal{F}(\mathcal{P},x)=\sum\limits_{q=0}^{\infty }\mathfrak{P}(q,1)x^{q}=%
\frac{h_{0}(\mathcal{P})+h_{1}(\mathcal{P})x}{(1-x)^{1+1}}  \tag{1.27}
\end{equation}%
so that Eq.(1.20) holds thus indicating presence of mirror symmetry. At this
point our readers might notice that, actually, for this symmetry to take
place one should consider instead of amplitude $A(s,t)$ the following
combination%
\begin{eqnarray}
A(s,t) &=&-g^{2}\frac{\Gamma (1-\alpha (s))\Gamma (1-\alpha (t))}{\Gamma
(1-\alpha (s)-\alpha (t))}+g^{2}\frac{\Gamma (-\alpha (s))\Gamma (-\alpha
(t))}{\Gamma (-\alpha (s)-\alpha (t))}  \notag \\
&=&-\hat{g}^{2}B(1-\alpha (s),1-\alpha (t))+g^{2}B(-\alpha (s),-\alpha (t)) 
\TCItag{1.27}
\end{eqnarray}%
Such a combination produces first two terms (with correct signs) of the
infinite series as proposed by Mandelstam, Eq.(15), Ref.[23]. The comparison
with experiment displayed in Fig.6.2 (a), Ref[22, p. 321] is quite
satisfactory producing one parent and one daughter trajectory. These are
also displayed in Ref.[20, p.41] for the "rho family". \ \ It should be
noted that in the present case the phase factors (eliminating tachyons)
discussed in Part I are not displayed since in Ref.[24] and, therefore also
in this work, we provide alternative explanation why tachyons should be
excluded from consideration.

\subsection{Organization of the rest of the paper}

In Section 2 using the concept of a zonotope we prove that, indeed,
Eq.(1.11) (and, hence, Eq.s (1.1) and (1.3)) is a special case of the Weyl
character formula. In arriving at this result we employ some information
about the Ruelle's dynamical transfer operator and earlier results by Atiyah
and Bott on Lefshetz-type fixed point formula for the elliptic complexes.
Section 3 along with results of Appendix (Part II) provides necessary
mathematical background to be used in Section 4. It includes some relevant
facts from the theory of toric varieties, algebraic groups, semisimple Lie
groups and assiciated with them Lie algebras, flag decompositions, etc. With
help of this information in Section 4 major physical applications are
developed culminating in the exact symplectic solution of the Veneziano
model. Connection between the symplectic and supersymmetric formalisms was
noticed and developed in the calssical paper by Atiyah and Bott, Ref.[25],
whose work had been inspired by earlier result by Witten, Ref. [26], on
supersymmetry and Morse theory. Thus, in view of Ref.[25], the results of
Part II and III become interrelated. The final results obtained in this work
\ are are in formal accord with those obtained earlier by Vergne, Ref.[3],
by other methods.

\section{From \ geometric progression to Weyl character formula}

\subsection{From p-cubes to d-polytopes via zonotope construction}

In Section 1 we have obtained Eq.(1.11). In one dimensional case it is
formally reduced to a simple geometric progression formula. The result,
Eq.(1.11), is obtained for the rational (or integral) polytope of cubic
symmetry. In this subsection we generalize this result to obtain similar
results for rational centrally symmetric polytopes whose symmetry is other
than cubic. This is possible with help of the concept of zonotope. The
concept of zonotope is not new. According to Coxeter [27\textbf{]}, it
belongs to the 19th century Russian crystallographer Fedorov. Nevertheless,
this concept has been truly appreciated only relatively recently in
connection with oriented matroids. For the purposes of this work it is
sufficient to consider only the most elementary properties of zonotopes.
Thus, following Ref.[28], let us consider a $p-$dimensional cube $C_{p}$
defined by 
\begin{equation}
C_{p}=\{\mathbf{x}\in \mathbf{R}^{p},-1\leq x_{i}\leq 1,\text{ }i=1-p\} 
\tag{2.1}
\end{equation}%
and the \ surjective map $\pi :\mathbf{R}^{p}\rightarrow \mathbf{R}^{d}$ .
The map is defined via the following

\textbf{Definition 2.1}. A \textit{zonotope} $\ Z(V)$ is the image of a $p-$%
cube, Eq.(2.1), under the affine projection $\pi.$ Specifically, 
\begin{align*}
Z & \equiv Z(V)=\mathbf{V}C_{p}+\mathbf{z}\ =\{V\mathbf{y}+\mathbf{z:y\in }%
C_{p}\} \\
& =\{x\in\mathbf{R}^{d}:x=z+\sum\limits_{i=1}^{p}x_{i}\mathbf{v}_{i}\text{ }%
,-1\leq x\leq1\}
\end{align*}
for some matrix (vector configuration) $\mathbf{V}=\{\mathbf{v}_{1},...,%
\mathbf{v}_{p}\}.$

By construction, such an image is a centrosymmetric $d$-polytope [28].
Below, we shall obtain some results for these $d$-polytopes. By
construction, they should hold also for $p$-cubes. In such a way we shall
demonstrate that, indeed, Eq.(1.11) can be associated with the Weyl
character formula.

\subsection{From Ruelle dynamical transfer operator to Atiyah and Bott
Lefschetz-type fixed point formula for the elliptic complexes}

Any classical dynamical system can be thought of as the pair $(\mathcal{M},f)
$ with $f$ being a map $f$: $\mathcal{M}\rightarrow \mathcal{M}$ from the
phase space $\mathcal{M}$ to itself. Following Ruelle[29], we consider a map 
$f$: $\mathcal{M}\rightarrow \mathcal{M}$ and a scalar function (a weight
function) $g$: $\mathcal{M}\rightarrow \mathbf{C.}$ Based on these data, the
transfer operator $\mathcal{L}$ can be defined as follows: 
\begin{equation}
\mathcal{L}\Phi (x)=\sum\limits_{y:fy=x}g(y)\Phi (y).  \tag{2.2}
\end{equation}%
If \ $\mathcal{L}_{1}$ and $\mathcal{L}_{2}$ are two such transfer operators
associated with some successive maps $f_{1}$ , $f_{2}$ : $\mathcal{M}%
\rightarrow \mathcal{M}$ and weights $g_{1}$ and $g_{2}$ then, 
\begin{equation}
(\mathcal{L}_{1}\mathcal{L}_{2}\Phi
)(x)=\sum\limits_{y:f_{2}f_{1y}=x}g_{2}(f_{1}y)g_{1}(y)\Phi (y).  \tag{2.3}
\end{equation}%
\ It is possible to demonstrate that 
\begin{equation}
tr\mathcal{L}=\sum\limits_{x\in Fixf}\frac{g(x)}{\left\vert \det
(1-D_{x}f^{-1}(x))\right\vert }  \tag{2.4}
\end{equation}%
with $D_{x}f$ being derivative of $f$ acting in the tangent space $T_{x}%
\mathcal{M}$ and the graph of $f$ is required to be transversal to the
diagonal $\Delta \subset \mathcal{M}\times \mathcal{M}$ . Eq.(2.4) coincides
with that obtained \ in the work by Atiyah and Bott [30]\footnote{%
They use $D_{x}f$ instead of $D_{x}f^{-1}$ which makes no difference for the
fixed points and invertible functions. The \ important (for chaotic
dynamics) non invertible case is discussed by Ruelle also but the results
are not much different.}. \ In connection with Eq.(2.4), these authors make
several important\ (for purposes of this work) observations to be discussed
now. In Ref.[29] Ruelle essentially uses the same type of arguments as those
by Atiyah and Bott [30]. These are given as follows. Define the \textit{local%
} Lefschetz index $\mathcal{L}_{x}(f)$ by 
\begin{equation}
\mathcal{L}_{x}(f)=\frac{\det (1-D_{x}f(x))}{\left\vert \det
(1-D_{x}f(x))\right\vert },  \tag{2.5}
\end{equation}%
where $x\in Fixf$. Then define the \textit{global} Lefschetz index $\mathcal{%
L(}\mathit{f)}$ by 
\begin{equation}
\mathcal{L(}\mathit{f)=}\sum\limits_{f(x)=x}\mathcal{L}_{x}(f).  \tag{2.6}
\end{equation}%
Taking into account that $\det (1-D_{x}f(x))$=$\prod\limits_{i=1}^{d}(1-%
\alpha _{i}),$ where $\alpha _{i\text{ }}$are the eigenvalues of the
Jacobian matrix, the determinant can be rewritten in the following useful
form, Ref.[31, p.133], 
\begin{equation}
\det (1-D_{x}f(x))=\prod\limits_{i=1}^{d}(1-\alpha
_{i})=\sum\limits_{k=0}^{d}(-1)^{k}e_{k}(\alpha _{1},...,\alpha _{d}), 
\tag{2.7}
\end{equation}%
where the elementary symmetric polynomial $e_{k}(\alpha _{1},...,\alpha _{d})
$ \ is defined by 
\begin{equation}
e_{k}(\alpha _{1},...,\alpha _{d})=\sum\limits_{1\leq \text{ }i_{1}<\cdot
\cdot \cdot <\text{ }i_{k}\leq d}\alpha _{i_{1}}\cdot \cdot \cdot \alpha
_{i_{k}}  \tag{2.8}
\end{equation}%
with $e_{k=0}=1.$With help of these results the local Lefschetz index,
Eq.(2.5), can be rewritten \ alternatively as follows 
\begin{equation}
\mathcal{L}_{x}(f)=\frac{\sum\limits_{k=01}^{d}(-1)^{k}e_{k}(\alpha
_{1},...,\alpha _{d})}{\left\vert \det (1-D_{x}f(x))\right\vert }\equiv 
\frac{\sum\limits_{k=0}^{d}(-1)^{k}tr(\wedge ^{k}D_{x}f(x))}{\left\vert \det
(1-D_{x}f(x))\right\vert }  \tag{2.9}
\end{equation}%
with $\wedge ^{k}$denoting \ the $k$-th power of the exterior product. Using
this result, Ruelle in Ref.[29] defined additional transfer operator $%
\mathcal{L}^{(k)}$ (analogous to earlier introduced $\mathcal{L)}$ as
follows. 
\begin{equation}
tr\mathcal{L}^{(k)}=\sum\limits_{x\in Fixf}\frac{g(x)tr(\wedge ^{k}D_{x}f(x))%
}{\left\vert \det (1-D_{x}f^{-1}(x))\right\vert }.  \tag{2.10}
\end{equation}%
In view of \ Eq.s(2.5)-(2.10), we also obtain, 
\begin{equation}
\sum\limits_{k=0}^{d}(-1)^{k}tr\mathcal{L}^{(k)}=\sum\limits_{x\in Fixf}g(x)%
\mathcal{L}_{x}(f).  \tag{2.11}
\end{equation}%
If in the above formulas we replace \ $Fixf$ by $Fixf^{n},$\ \ we have to
replace $tr\mathcal{L}^{(k)}$ by $tr\mathcal{L}_{n}^{(k)}$. Next, since 
\begin{equation}
\exp (\sum\limits_{n=1}^{\infty }\frac{tr(\mathbf{A}^{n})}{n}t^{n})=\left[
\det (\mathbf{1}-t\mathbf{A})\right] ^{-1}  \tag{2.12}
\end{equation}%
it is convenient to combine this result with Eq.(2.10) in order to obtain
the following dynamical zeta function: 
\begin{align}
Z(t)& =\exp (\sum\limits_{n=1}^{\infty }\frac{t^{n}}{n}\left\{
\sum\limits_{k=0}^{d}(-1)^{k}tr\mathcal{L}_{n}^{(k)}\right\} )  \notag \\
& =\prod\limits_{k=0}^{d}\left[ \exp (\sum\limits_{n=1}^{\infty }\frac{tr%
\mathcal{L}_{n}^{(k)}}{n}t^{n})\right] ^{(-1)^{k}}  \notag \\
& =\prod\limits_{k=0}^{d}\left[ \det (\mathbf{1-}t\mathcal{L}^{(k)})\right]
^{(-1)^{k+1}}.  \tag{2.13}
\end{align}%
This final result coincides with that obtained by Ruelle as required. Thus
obtained zeta function possess dynamical, number-theoretic and
algebro-geometric interpretation. Looking at Eq.(1.1), it should be clear
that for the appropriately chosen $d$ and $\mathcal{L}$ Eq.(1.1) and (2.13)
can be made the same. Moreover, based on the paper by Atiyah and Bott,
Ref.[30], we would like to demonstrate that Eq.(2.4) is actually the same
thing as the Weyl's character formula [32]. To prove that this is indeed the
case is not entirely trivial. In what follows, we shall assume that our
readers are familiar with results and notations of Part II and, especially,
with the results of Appendix to Part II. To avoid duplications, we shall use
below results on the Weyl-Coxeter reflection groups using notations from
this Appendix\footnote{%
To shorten notations we shall write "Appendix" having in mind the Appendix
of Part II.}

\subsection{From Atiyah-Bott-Lefschetz fixed point formula to character
formula by Weyl}

We begin with observation that Eq.s(2.4) and (1.11) are equivalent. Because
of this, it is sufficient to demonstrate that the r.h.s of Eq.(1.11) \
indeed coincides with the Weyl's character formula. Although Eq.(1.11) (and,
especially, Eq.s(2.3) and (2.10)) looks similar to that obtained in the
paper by Atiyah and Bott (AB), Ref.[30, part I, p.379], leading to the Weyl
character formula, Eq.(5.12) of Ref.[30\textbf{,} part II\footnote{%
It should be takent into account that the AB paper aslo has Parts I and II.}%
], neither Eq.(1.11) nor Eq.(5.11) of AB paper Ref.[30, part II] provide
immediate connection with their Eq.(5.12). Hence, the task now is to restore
some missing links.

To this purpose we need to recall some facts from the book by Bourbaki,
Ref.[32].These facts are also helpful in the remainder of this paper. In
particular, let us consider a finite set of formal symbols $e(\mu )$
possessing the same multiplication properties as the usual exponents%
\footnote{%
In the case of \textit{usual} exponents it is being assumed that all the
properties of formal exponents are transferable to the usual ones.}, i.e. 
\begin{equation}
e(\mu )e(\nu )=e(\mu +\nu ),\text{ }\left[ e(\mu )\right] ^{-1}=e(-\mu )%
\text{ and }e(0)=1.  \tag{2.14}
\end{equation}%
Such defined set of formal exponents is making a free $\mathbf{Z}$ module
with the basis $e(\mu ).$ Subsequently, we shall require that $\mu \in
\Delta $ with $\Delta $ being the Weyl root system defined in the Appendix.
Suppose also that we are given a polynomial ring $A[\mathbf{X}]$ made of all
linear combinations of terms $\mathbf{X}^{\mathbf{n}}\equiv
X_{1}^{n_{1}}\cdot \cdot \cdot X_{d}^{n_{d}}$ with $n_{i}\in \mathbf{Z}$ and 
$X_{i}$ being some indeterminates. Then, one can construct another ring A[P]
made of linear combinations of elements $e(\mathbf{p}\cdot \mathbf{n})$ with 
$\mathbf{p}\cdot \mathbf{n=}p_{1}n_{1}+\cdot \cdot \cdot p_{d}n_{d}$.
Clearly, the above rings are isomorphic as it was explained in Part II,
Section 9. Let $x=\sum\limits_{p\in P}x_{p}e(p)\in $A[\textbf{P}] \ with 
\textbf{P}=$\{p_{1},...,p_{d}\}.$ Then using Eq.(2.14) we obtain, 
\begin{align}
x\cdot y& =\sum\limits_{s\in P}x_{s}e(s)\sum\limits_{r\in
P}y_{r}e(r)=\sum\limits_{t\in P}z_{t}e(t)\text{ with}  \notag \\
z_{t}& =\sum\limits_{s+r=t}x_{s}y_{r}\text{ and, accordingly,}  \notag \\
x^{m}& =\sum\limits_{t\in P}z_{t}e(t)\text{ with }z_{t}=\sum\limits_{s+\cdot
\cdot \cdot +r=t}x_{s}\cdot \cdot \cdot y_{r}\text{, }m\in \mathbf{N} 
\tag{2.15}
\end{align}%
with $\mathbf{N}$ being some non negative integer. Introduce now the
determinant of $w\in W$ via rule: 
\begin{equation}
\det (w)\equiv \varepsilon (w)=(-1)^{\mathit{l}(w)},  \tag{2.16}
\end{equation}%
where, again, we use notations from Appendix. If, in addition, we would
require 
\begin{equation}
w(e(p))=e(w(p)),  \tag{2.17}
\end{equation}%
then all elements of the ring A[\textbf{P}] are subdivided into two classes
defined by 
\begin{equation}
w(x)=x\text{ (invariance)}  \tag{2.18a}
\end{equation}%
and 
\begin{equation}
w(x)=\varepsilon (x)\cdot x\text{ (anti invariance).}  \tag{2.18b}
\end{equation}%
These classes are very much like subdivision into bosons and fermions in
quantum mechanics\footnote{%
Such analogy is not superficialas we have noticed already in Part II.}. All
anti invariant elements can be built from the basic anti invariant element $%
J(x)$ which, in view of Eq.(2.7), can defined by 
\begin{equation}
J(x)=\sum\limits_{w\in W}\varepsilon (w)\cdot w(x).  \tag{2.19}
\end{equation}%
From the definition of the set \textbf{P} and from Appendix it should be
clear that the set \textbf{P} can be identified with the set of reflection
elements $w$ of the Weyl group $W$. Therefore, for all $x\in $ A[\textbf{P}]
and $w\in W$ we obtain the following chain of equalities: 
\begin{equation}
w(J(x))=\sum\limits_{v\in W}\varepsilon (v)\cdot w(v(x))=\varepsilon
(w)\sum\limits_{v\in W}\varepsilon (v)\cdot v(x)=\varepsilon (w)J(x) 
\tag{2.20}
\end{equation}%
as required. Accordingly, \textit{any} anti invariant element $x$ can be
written as $x=$ $\sum\limits_{l\in P}x_{p}J(\exp (p)).$ \ The denominator of
Eq.(1.11), when \ properly interpreted with help of results of  Appendix,
can be associated with $J(x)$. Indeed, without loss of generality let us
choose the constant $\mathbf{c}$ as $\mathbf{c}=\{1,...,1\}$. Then, for the
fixed $v$ the denominator of Eq.(1.11) can be rewritten as follows: 
\begin{align}
\prod\limits_{i=1}^{d}(1-\exp \{-u_{i}^{v}\})& \equiv \prod\limits_{\alpha
\in \Delta ^{+}}(1-\exp (-\alpha ))\equiv \tilde{d}\exp (-\rho ),\text{ } 
\tag{2.21} \\
\text{where }\rho & =\frac{1}{2}\sum\limits_{\alpha \in \Delta ^{+}}\alpha 
\text{ and }  \notag \\
\tilde{d}& =\prod\limits_{\alpha \in \Delta ^{+}}(\exp (\frac{\alpha }{2}%
)-\exp (-\frac{\alpha }{2})).  \tag{2.22}
\end{align}%
To prove that $\ $thus defined $\tilde{d}$ belongs to the anti invariant
subset of A[P] is not difficult. Indeed, consider the action of a reflection 
$r_{i}$ on $\tilde{d}$. Taking into account that $r_{i}(\alpha _{i})=-\alpha
_{i}$ we obtain, 
\begin{align}
r_{i}(\tilde{d})& =(\exp (-\frac{\alpha _{i}}{2})-\exp (\frac{\alpha _{i}}{2}%
))\prod\limits_{\substack{ \alpha \neq \alpha _{i} \\ \alpha \in \Delta ^{+}
}}(\exp (\frac{\alpha }{2})-\exp (-\frac{\alpha }{2}))  \notag \\
& =-\tilde{d}\equiv \varepsilon (r_{i})\tilde{d}.  \tag{2.23}
\end{align}%
Hence, clearly, 
\begin{equation}
\tilde{d}=\sum\limits_{p\in P}x_{p}J(\exp (p)).  \tag{2.24}
\end{equation}%
Moreover, it can be shown [32], that $\tilde{d}=J(\exp (\rho ))$ which, in
view of Eq.s (2.21), (2.22), produces identity originally obtained by Weyl : 
\begin{equation}
\hat{d}\exp (-\rho )=\prod\limits_{\alpha \in \Delta ^{+}}(1-\exp (-\alpha
)).  \tag{2.25}
\end{equation}%
Applying reflection $w$ to the above identity while taking into account
Eq.s(2.17), (2.23) produces: 
\begin{equation}
\prod\limits_{\alpha \in \Delta ^{+}}(1-\exp (-w(\alpha )))=\exp (-w(\rho ))%
\text{ }\varepsilon (w)\text{ }\hat{d}.  \tag{2.26}
\end{equation}%
The result just obtained is of central importance for the proof of the
Weyl's formula. Indeed, in view of Eq.s (2.17) and (2.26), inserting \ the
identity : $1=\dfrac{w}{w}$ $\ $\ into the sum over the vertices on the
r.h.s. of Eq.(1.11) \ and taking into account that: a) $\varepsilon (w)=\pm 1
$ so that $\left[ \varepsilon (w)\right] ^{-1}=\varepsilon (w)$ , b)
actually, the sum over the vertices is the same thing as \ the sum \ over
the members of the Weyl-Coxeter group (since all vertices of the \ integral
polytope can be obtained by using the appropriate reflections applied to the
highest weight vector pointing to\ a chosen vertex), we obtain the Weyl's
character formula: 
\begin{equation}
tr\mathcal{L(\lambda )=}\frac{1}{\hat{d}}\sum\limits_{w\in \Delta
}\varepsilon (w)\text{ exp}\{w(\lambda +\rho )\}.  \tag{2.27}
\end{equation}%
It was obtained with help of the results of Appendix, Eq.s(2.4),(2.17) and
(2.26). Looking at the l.h.s. of Eq.(1.11) we can replace $tr\mathcal{%
L(\lambda )}$ by the sum in the l.h.s. of Eq.(1.11) if we choose the
constant $\mathbf{c}$ as before. \ Doing this is not too illuminating
however as we would like to explain now.

Indeed, since by construction $J(x)$ in Eq.(2.19) is the basic anti
invariant element and the r.h.s of Eq.(2.27) is by design manifestly
invariant element of the A[\textbf{P}], it is only natural to look for the
basic invariant element analogue of $J(x).$ Then, in view of Eq.(2.15) (and
discussion preceding this equation), $tr\mathcal{L(\lambda )\equiv \chi
(\lambda )}$ should be expressible as follows: 
\begin{equation}
\mathcal{\chi (\lambda )=}\sum\limits_{w\in W}n_{w}(\lambda )\text{ }e(w). 
\tag{2.28}
\end{equation}%
The factor $n_{w}(\lambda )$ in Eq.(2.28) is known in group theory as the
Kostant's multiplicity formula [33]. It plays an important role in our work,
especially in Section 4. To calculate it explicitly, Cartier [34] simplified
the original derivation by Kostant. In view of simplicity of his arguments,
we would like to reproduce them having in mind their later use in the text.
Cartier noticed that the denominator of the Weyl character formula,
Eq.(2.27), can be formally expanded with help of Eq.(2.25) as follows: 
\begin{equation}
\left[ \exp (\rho )\prod\limits_{\alpha \in \Delta ^{+}}(1-\exp (-\alpha ))%
\right] ^{-1}=\sum\limits_{w^{\prime }\in W}P(w^{\prime })e(-\rho -w^{\prime
}).  \tag{2.29}
\end{equation}%
By combining Eq.s(2.27),(2.28) and (2.29) we obtain, 
\begin{equation}
\sum\limits_{w\in W}n_{w}(\lambda )\text{ }e(w)=\sum\limits_{w\in
W}\varepsilon (w)\text{ exp}\{w(\lambda +\rho )\}\sum\limits_{w^{\prime }\in
W}P(w^{\prime })e(-\rho -w^{\prime }).  \tag{2.30}
\end{equation}%
Comparing the l.h.s. with the r.h.s. in the above expression we obtain
finally the Kostant multiplicity formula: 
\begin{equation}
n_{w}(\lambda )=\sum\limits_{w^{\prime }\in W}\varepsilon (w^{\prime
})P(w^{\prime }(\lambda +\rho )-(\rho +w)).  \tag{2.31}
\end{equation}%
The obtained formula allows us to determine the density of states $%
n_{w}(\lambda )$ Provided that the function $P$ is known explicitly, the
obtained formula allows us to determine the factor $n_{w}(\lambda )$.

It is useful to rewrite these results in physical language. In particular,
for any quantum mechanical system, the partition function $\ \Xi $ can be
written as 
\begin{equation}
\Xi =\sum\limits_{n}g_{n}\exp \{-\beta E_{n}\}\equiv tr(\exp (-\beta \hat{H}%
))  \tag{2.32}
\end{equation}%
where $\hat{H}$ is \ the quantum Hamiltonian of the system, $\beta $ is the
inverse temperature and $g_{n}$ is the degeneracy factor. Clearly, using
Eq.s(2.27), (2.28), one can identify $\Xi $ with $\mathcal{\chi (\lambda )}$%
. Next we introduce the density of states $\rho (E)$ via 
\begin{equation}
\rho (E)=\sum\limits_{n}g_{n}\delta (E-E_{n}).  \tag{2.33}
\end{equation}%
Comparison between Eq.(2.31) and (2.33) \ suggests that the function $P$ can
be identified with the density of states. Using $\rho (E)$ the partition
function $\Xi $ can be written as the Laplace transform 
\begin{equation}
\Xi (\beta )=\int\limits_{0}^{\infty }dE\rho (E)\exp \{-\beta E\}. 
\tag{2.34}
\end{equation}%
Clearly, Eq.(2.28) is just the discrete analogue of Eq.(2.34) so that it
does have a statistical/quantum mechanical interpretation as partition
function. From condensed matter physics it is known that all important
statistical/quantum information is contained in the density of states. Its
calculation is of primary interest in physics. Evidently, the same is true
in the present case.

In the light of results just obtained, we can reinterpret some results from
the Introduction. In particular, the r.h.s. our Eq.(1.11), when compared
with the r.h.s. of Eq.(5.11) of AB paper, Ref.[30, part II], is not looking
the same. We would like to explain that, nevertheless, these expressions are
equivalent. For comparison, let us reproduce Eq.(5.11) of AB paper first.
Actually, for this purpose it is more convenient to use the paper by Bott,
Ref.[35], (his Eq.(28)).  In notations taken from this reference, Eq.(5.11)
by AB is written now as follows: 
\begin{equation}
trace\text{ }T_{g}=\sum\limits_{\substack{ w\in W \\ \alpha <0}}\left[ \frac{%
\lambda }{\prod \left( 1-\alpha \right) }\right] ^{w}.  \tag{2.35}
\end{equation}%
Comparing \ this result with the r.h.s. of our Eq.(1.11) and taking into
account Eq.(2.17), the combination $\lambda ^{w}$ in the numerator of
Eq.(2.35) is the same thing as $\exp \{w\lambda \}$ in Eq.(2.27). As for the
denominator, Bott uses the same Eq.(2.25) as we do so that it remains to
demonstrate that 
\begin{equation}
\left[ \prod\limits_{\alpha \in \Delta ^{+}}(1-\exp (-\alpha ))\right]
^{w}=\exp (-w(\rho ))\text{ }\varepsilon (w)\text{ }\hat{d}.  \tag{2.36}
\end{equation}%
In view of Eq.(2.25), we need to demonstrate that 
\begin{equation}
\left[ \hat{d}\exp (-\rho )\right] ^{w}=\exp (-w(\rho ))\text{ }\varepsilon
(w)\text{ }\hat{d},  \tag{2.37}
\end{equation}%
i.e. that $\left[ \hat{d}\right] ^{w}=\varepsilon (w)$ $\hat{d}.$ Looking at
Eq.(2.23), this requires us to assume that $\left[ \hat{d}\right] ^{w}=w\hat{%
d}.$ But, in view of Eq.s(2.17), (2.19) and (2.24), we conclude that this is
indeed the case. This proves the fact that Eq.(2.35), that is Eq.(5.11) of
Ref.[30], is indeed the same thing as the Weyl's character formula,
Eq.(5.12) of Ref.[30], or, equivalently, Eq.(2.27) above. According to Kac,
Ref.[36, p.174], the classical Weyl character formula, Eq.(2.27), is
formally valid for both finite dimensional semisimple Lie algebras and
infinite dimensional affine Kac-Moody algebras. This circumstance and the
Proposition A.1. of Appendix play important motivating role for developments
in our work.

\textbf{Remark 2.2. }Although our arguments thus far have been limited only
to (comparison with) the $d-$ dimensional hypercube, this deficiency is
easily correctable with help of the concept of \textit{zonotope \ }%
introduced in Section 2.1. Clearly, \ because of zonotope construction the
obtained results remain correct for \textit{any} centrally symmetric
polytope $\mathcal{P}$. Thus, we have demonstrated that Eq.(1.11) has
essentially the same meaning as the Weyl character formula.

\textbf{Remark 2.3.} In view of Eq.s(2.22),(2.35) the denominator $\hat{d}$
in the Weyl character formula, Eq.(2.27), is actually a determinant. This
means that the basic anti invariant $J(x)$ introduced in Eq.(2.19) is
determinant. But the r.h.s. of Eq.(2.27) is invariant. This is possible only
if the numerator of the Weyl character formula is also a determinant. Hence,
the Weyl character formula is essentially the ratio of determinants. Since
this is surely the case, it implies that the Ruele zeta function, Eq.(2.13)
is also essentially the Weyl character formula\footnote{%
That this is the case for dynamical systems can be deduced, based on our
arguments, from the monograph by Feres, Ref.[37\textbf{]}. We shall not
develop this line of thought in this paper having in mind different goals.}.
This, in turn, implies that our major result for the Veneziano partition
function, Eq.(1.1), is essentially also the Weyl character formula in accord
with Lerche et al [4], where such conclusion was obtained differently.

The fact that Eq.(1.1) can be interpreted as the Weyl character formula
should not be too surprising in view of the fact that the l.h.s. of Eq.(1.1)
denotes the Poincare polynomial which is group invariant. It remains to
demonstrate that the torus action group introduced in Part II can be
interpreted as the Weyl-Coxeter reflection group. This is done below, in
Section 3. In the meantime, we have not exhausted all consequences of the
results we have just obtained. In particular, if it is true that Eq.(1.11)
is the Weyl character formula, then, taking into account Eq.(2.28), we
conclude that Eq.(1.7) is the Kostant multiplicity formula for $d$%
-dimensional cube. The rigorous proof of this fact for the arbitrary convex
polytope can be found in Refs. [33, 38, 39]. If this is so, then Eq.(1.2) is
also the Kostant multiplicity formula providing the number of points inside
the inflated (with inflation factor $q$) simplex $q\Delta _{n}$ (living in 
\textbf{Z}$^{n}$ lattice) and at its boundaries. These observations allow us
to develop symplectic methods for reconstruction of the Veneziano partition
function to be discussed below in Section 4.

\textbf{Remark 2.4}. From the point of view of algebraic geometry of toric
varieties [40-42], the Kostant multiplicity formula has yet another
(topological) interpretation as the Euler characteristic $\chi $ of the 
\textit{projective} toric variety associated with the polytope $\mathcal{P}$%
. This fact will be discussed in some detail \ below, in Sections 3.4 and 4.
It \ motivates us to develop symplectic formulation of the Veneziano
partition function in Section 4 and supersymmetric formulation discussed in
Part II. Connections between the Weyl character formula and the Euler
characteristic for projective algebraic varieties had been uncovered by
Nielsen, Ref.[43], already in 1976. His work is based on still earlier work
by Iversen and Nielsen, Ref.[44]. Below we shall argue that, actually, the
idea of such connection can be traced back to still much earlier papers by
Hopf, Ref.[45], and Hopf and Samelson, Ref.[46]. In Section 3.4., we provide
the topological interpretation of the Kostant multiplicity formula in terms
of $\chi $ based on ideas of Hopf and Samelson.

\bigskip

\section{ The torus action and the moment map}

\subsection{ The torus action and the Weyl group}

\bigskip

To avoid duplications, in writing this and following subsections we shall
assume that our readers are familiar with our earlier work, Part II. Hence,
in this part we only introduce terminology which is of immediate use. To
begin, let us consider a polynomial 
\begin{equation}
f(\mathbf{z})=f(z_{1},...,z_{n})=\sum\limits_{\mathbf{i}}\lambda_{\mathbf{i}}%
\mathbf{z}^{\mathbf{i}}=\sum\limits_{\mathbf{i}%
}\lambda_{i_{1}....i_{n}}z_{1}^{i_{1}}\cdot\cdot\cdot z_{n}^{i_{n}},\text{ (}%
\lambda_{\mathbf{i}},\text{ }z_{m}^{i_{m}}\in\mathbf{C,}1\leq m\leq n). 
\tag{3.1}
\end{equation}
It belongs to the polynomial ring $A[\mathbf{z}]$ (essentially isomorphic to
earlier introduced $A[\mathbf{X}])$ closed under ordinary addition and
multiplication. Since now we are using complex numbers (instead of
indeterminates as in Section 2.3.) this allows us to introduce the following

\textbf{Definition 3.1}.\ An\textit{\ affine} algebraic variety $V\in 
\mathbf{C}^{n}$ is the set of zeros of the\ collection of polynomials from
the ring $A[\mathbf{z}].$ \ \ \ \ \ \ \ \ \ \ \ \ \ \ \ \ \ \ \ \ \ \ \ \ 

According to the famous Hilbert's Nullstellensatz a collection of such
polynomials is \textit{finite }and forms the set \ $I(\mathbf{z}):=\{f\in $
A[$\mathbf{z}$]$,$ $f(\mathbf{z})=0\}$ of maximal ideals usually denoted 
\emph{Spec}A[\textbf{z}].

\textbf{Definition 3.2.}The zero set of a \textit{single} function belonging
to $I(\mathbf{z})$ is called \textit{algebraic hypersurface }so that the
set\ $I(\mathbf{z})$ corresponds to the \textit{intersection} of a finite
number of hypersurfaces.

As in Part II, we need to consider the set of Laurent monomials of the type $%
\lambda \mathbf{z}^{\mathbf{\alpha }}\equiv \lambda z_{1}^{\alpha _{1}}\cdot
\cdot \cdot z_{n}^{\alpha _{n}}$ . We shall be particularly interested in
the \textit{monic} monomials for which $\lambda =1.$ Such monomials form a
closed polynomial \textit{subring} with respect to usual multiplication and
addition. The crucial step forward is to assume that the exponent $\mathbf{%
\alpha \in S}_{\sigma }\footnote{%
Although the monoid $\mathbf{S}_{\sigma }$ was defined in Part II, for
reader's convenience it will be redefined below momentarily}$. This allows
us to define the following mapping 
\begin{equation}
u_{i}:=z^{a_{i}}  \tag{3.2}
\end{equation}%
with $a_{i}$ being \ one of the generators of the monoid $\mathbf{S}_{\sigma
}$ and$\ z\in \mathbf{C}.$ In order to define the monoid $\mathbf{S}_{\sigma
}$ we still need to provide a couple of definitions. In particular, recall
that \textit{a semi-group }$\mathit{S}$\textit{\ }that is a non-empty set
with associative\textit{\ }operation\textit{\ }is called \textit{monoid\ }if
it is commutative, satisfies cancellation law\textit{\ (i.e.s+x=t+x }implies%
\textit{\ s=t for \ all s,t,x}$\in S)$ and has zero element $(i.e.s+0=s,s\in
S).$This allows us to make the following

\textbf{Definition 3.3}. \textit{A monoid }$\mathit{S}$\textit{\ is} \textit{%
finitely generated }if exist a set\textit{\ }$a_{1},...,a_{k}\in S$\textit{, 
}called $generators$, such that 
\begin{equation}
S=Z_{\geq 0}a_{1}+\cdot \cdot \cdot +Z_{\geq 0}a_{k}.  \tag{3.3}
\end{equation}%
Taking into account this definition, it is clear that the monoid $\mathbf{S}%
_{\sigma }$ =$\sigma \cap \mathbf{Z}^{d}$ \ for the rational polyhedral cone 
$\sigma $(e.g. read Part II) is \textit{finitely generated}.

\ \ \ \ The mapping given by Eq.(3.2) provides an isomorphism between the 
\textit{additive} group of exponents $a_{i}$ and the \textit{multiplicative}
group of monic Laurent polynomials. Next, the function $\phi $ is considered
to be \textit{quasi homogenous} of degree $d$ with exponents \textit{l}$%
_{1},...,l_{n}$ if 
\begin{equation}
\phi (\lambda ^{\mathit{l}_{1}}x_{1},...,\lambda ^{\mathit{l}%
_{n}}x_{n})=\lambda ^{d}\phi (x_{1},...,x_{n}),  \tag{3.4}
\end{equation}%
provided that $\lambda \in \mathbf{C}^{\ast }.$ Applying this result to $z^{%
\mathbf{a}}\equiv z_{1}^{a_{1}}\cdot \cdot \cdot z_{n}^{a_{n}}$ we obtain
equation\ analogous to Eq.(3.3) for the polyhedral cone: 
\begin{equation}
\sum\limits_{j}\left( l_{j}\right) _{i}a_{j}=d_{i}.  \tag{3.5}
\end{equation}%
Clearly, if the index $i$ is numbering different monomials, then \ the sum $%
d_{i}$ belongs to the monoid $\mathbf{S}_{\sigma }.$ The same result can be
achieved if instead we would consider the products of the type $%
u_{1}^{l_{1}}\cdot \cdot \cdot u_{n}^{l_{n}}$ and rescale all $z_{i}^{\prime
}s$ by the same factor $\lambda .$ Eq.(3.5) should be understood as a scalar
product with $\left( l_{j}\right) _{i}$ living in the space \textit{dual} to 
$a_{j}^{\prime }s$ . Accordingly, the set of $\left( l_{j}\right)
_{i}^{\prime }s$ can be considered as the set of generators for the dual
cone $\sigma ^{\vee }.$ Next, in view of Eq.(3.2), let us consider the
polynomials of the type 
\begin{equation}
f(\mathbf{z})=\sum\limits_{\mathbf{a\in S}_{\sigma }}\lambda _{\mathbf{a}}%
\mathbf{z}^{\mathbf{a}}=\sum\limits_{\mathbf{l}}\lambda _{\mathbf{l}}\mathbf{%
u}^{\mathbf{l}}.  \tag{3.6}
\end{equation}%
As before, they form a polynomial ring. The ideal for this ring is
constructed based on the observation that for the fixed\ $d_{i}$ \ and the
assigned set of \ cone generators $a_{i}$ there is more than one set of
generators for the dual cone. This redundancy produces relations of the type 
\begin{equation}
u_{1}^{l_{1}}\cdot \cdot \cdot u_{k}^{l_{k}}=u_{1}^{\tilde{l}_{1}}\cdot
\cdot \cdot u_{k}^{\tilde{l}_{k}}.  \tag{3.7}
\end{equation}%
If now we require $u_{i}\in \mathbf{C}_{i},$ then it is clear that the above
equation belongs to the ideal $I(\mathbf{z})$ of the above polynomial ring
and that Eq.(3.7) represents the hypersurface in accord with the Definitions
3.1 and 3.2. \ As before, the ideal $I(\mathbf{z})$ represents the
intersection of these hypersurfaces thus forming the affine \textit{toric}
variety $X_{\sigma ^{\vee }}.$ The generators \{$u_{1},...,u_{k}\}\in 
\mathbf{C}^{k}$ are coordinates for $X_{\sigma ^{\vee }}$. They represent
the same point in $X_{\sigma ^{\vee }}$ if and only if \textbf{u}$^{\mathbf{l%
}}=$\textbf{u}$^{\mathbf{\tilde{l}}}.$ \ Thus formed toric variety
corresponds to just one (dual) cone. A \textit{complex algebraic torus} $T$
\ is defined by the rule: $T:=(\mathbf{C}\backslash 0)^{n}=:(\mathbf{C}%
^{\ast })^{n}.$It acts on the affine toric variety $X_{\Sigma }$ (made out
of pieces $X_{\sigma ^{\vee }}$ with help of a gluing map) according to the
rule : $T\times X_{\Sigma }\rightarrow X_{\Sigma }$, provided that at each
affine variety corresponding to the dual cone $\sigma ^{\vee }$ its action
is given by 
\begin{equation}
T\times X_{\sigma ^{\vee }}\rightarrow X_{\sigma ^{\vee }}\text{ , }%
(t,x)\mapsto tx:=(t^{a_{1}}x_{1},...,t^{a_{k}}x_{k}).  \tag{3.8}
\end{equation}

To proceed, we replace temporarily $T$ by the group $G$ acting
(multiplicatively) on the set $X$ via rule: $G\times X\rightarrow X,$ i.e. $%
(g,x)\rightarrow gx,$ provided that for all $g,h\in G$, $g(hx)=(gh)x$ and $%
ex=x$ for some unit element $e$ of $G$.

\bigskip

\textbf{Definition 3.4}. The subset $Gx:=\{gx\mid g\in X\}$ of $X$ is called
the \textit{orbit }of $x$. The subgroup $H:=\{gx=x\mid g\in X\}$ of $G$ that
fixes $x$ is the $isotropy$ group. There could be more than one fixed point
for the equation $gx=x$. All of them are conjugate to each other.

\textbf{Definition 3.5}. A \textit{homogenous} space for $G$ is the subspace
of $X$ on which $G$ acts without fixed points.

\bigskip

The major step forward can be made by introducing the concept of an \textit{%
algebraic} group \textbf{[}47].

\textbf{Definition 3.6.} A \textit{linear algebraic group} $G$ \ is a) an
affine algebraic variety and b) a group in the sense given above, i.e. 
\begin{align}
\mu & :G\times G\rightarrow G\text{ };\text{ }\mu (x,y)=xy  \tag{3.9a} \\
i& :G\rightarrow G\text{ };\iota (x)=x^{-1}.  \tag{3.9b}
\end{align}

\textbf{Remark.3.7}. It can be shown, Ref. [48, p.150], that $G$ as a linear
algebraic group is isomorphic to a closed subgroup of $GL_{n}($K$)$ for some 
$n\geq 1$ and any closed number field K such as $\mathbf{C}$ or $p-$adic.
This fact plays the central role in whole development presented below.

Consider therefore an action of $G$ on $f(\mathbf{z})$ defined by Eq.(3.6).
Following Stanley, Ref. [13], it can be defined as $M\circ f(\mathbf{z})=f(M%
\mathbf{z})$ for some matrix $M$ such that $M\in G.$ In order for this
definition to be compatible with earlier made, Eq.(3.8), we have to assume
that the torus $T$ acts diagonally on the vector space spanned by $%
x_{1},...,x_{n}.$ This means that the isotropy group of the torus is defined
by the set of the following equations 
\begin{equation}
t^{a_{i}}x_{i}=x_{i}\text{ .}  \tag{3.10}
\end{equation}%
Apart from trivial solutions$:x_{i}=0$ and $x_{i}=\infty $, there are
nontrivial solutions: $t^{a_{i}}=1$ $\forall $ $x_{i}$. For integer $%
a_{i}^{\prime }s$ these are cyclotomic equations whose $a_{i}-1$ solutions
all lie on the circle $S^{^{1}}($e.g. see PartI, Section 3.1.$)$ This result
is easy to understand since the algebraic torus $T$ has the topological
torus as a deformation retract while the topological torus is just a
Cartesian product of circles.

Next, we notice that Eq.(3.8) still makes sense if some of $t-$factors are
replaced by 1's. This means that one should take into account situations
when one, two, etc. $t$-factors in Eq.(3.8) are replaced by 1's and account
for all permutations involving such cases. This observation leads to the
torus actions on toric subvarieties. It is important that different orbits
belong to different subvarieties which do not overlap. Thus, by design, $%
X_{\Sigma }$ is the disjoint union of \textit{finite} number of orbits
identified with the subvarieties of $X_{\Sigma }.$Under such circumstances
the\ vector $\left( x_{1},...,x_{k}\right) $ forms a basis of \ $k-$%
dimensional vector space $V$ so that the vector $\left(
x_{1},...,x_{i}\right) $ , $i\leq k,$ forms a basis of the subspace $V_{i}$
. This allows us to introduce a complete flag $\ f_{0}$ of subspaces in $V$
via 
\begin{equation}
f_{0}:0=V_{0}\subset V_{1}\subset ...\subset V_{k}=V.  \tag{3.11}
\end{equation}%
Consider now an action of $G$ on $f_{0}$ . Taking into account the Remark
3.8., we recognize that effectively $G$=$GL_{n}(K)$. The matrix
representation of this group possess remarkable properties. These are
summarized in the following definitions.

\textbf{Definition 3.8. }Given that $\ $the set $GL_{n}(K)=\{x\in M_{n}(K$)$%
\mid \det x\neq 0\}$ with $M_{n}(K$) being \ $n\times n$ matrix with entries 
$x_{i,j}\in K$ forms a general linear group, the matrix $x\in M_{n}($K) is
a) $semisimple$ \ ($x=x_{s}$), if it is diagonalizable, that is $\exists
g\in GL_{n}(K)$ such that $gxg^{-1}$ is a diagonal matrix; b) $nilpotent$ ($%
x=x_{n}$ ) if $x^{m}=0,$that is for some positive integer $m$ all
eigenvalues of the matrix $x^{m}$ are zero; c) $unipotent$ ($x=x_{u}$), if $%
x-1_{n}$ is nilpotent, i.e. $x$ is the matrix whose only eigenvalues are 1's.

Just like with the odd and even numbers the above matrices, \textit{if they
exist,} form closed disjoint subsets of $GL_{n}(K)$, e.g. all $x,y\in
M_{n}(K)$ commute; if $x,y$ are semisimple so is their sum and the product,
etc. Most important for us is the following fact:

\bigskip

\textbf{Proposition 3.9}. \textit{Let }$x\in GL_{n}($\textit{K}$)$\textit{.
Then }$\exists $\textit{\ }$x_{u}$\textit{\ and }$x_{s}$\textit{\ such that }%
$x=x_{s}x_{u}=x_{u}x_{s}$\textit{\ Both x}$_{s}$\textit{and x}$_{u}$\textit{%
\ are determined by the above conditions uniquely.}

\textit{\bigskip}

The proof can be found in Ref.[49, p.96]. This proposition is in fact a
corollary of the Lie-Kolchin theorem \ which is of major importance for us.
To formulate this theorem \ we need to introduce yet another couple of
definitions. In particular, if $A$ and $B$ are closed (finite) subgroups of
the algebraic group $G$ one can construct the group $(A,B)$ made of
commutators $xyx^{-1}y^{-1}$, $x\in A,$ $y\in B$ . With help of such
commutators the following definition can be made

\textbf{Definition 3.10.} The group $G$ is \textit{solvable} if its derived
series terminates in the unit element $e$. The derived series is being
defined inductively by $\mathcal{D}^{\left( 0\right) }G=G,\mathcal{D}%
^{\left( i+1\right) }G=(\mathcal{D}^{\left( i\right) }G,\mathcal{D}^{\left(
i\right) }G),$ $i\geq 0.$

Such a definition implies that the algebraic group $G$ is solvable if and
only if there exists a chain $G=G^{\left( 0\right) }\supset G^{(1)}\supset
\cdot \cdot \cdot G^{\left( n\right) }=e$ for which $(G^{\left( i\right)
},G^{\left( i\right) })\subset G^{i+1}$ $(0\leq i\leq n)$, Ref.$[49$, p.111]$%
.$ Finally,

\textbf{Definition 3.11}.The group is called \textit{nilpotent} if $\mathcal{%
E}^{(n)}\mathit{G=e}$ for some $n$, where $\mathcal{E}^{\left( 0\right) }=G,%
\mathcal{E}^{\left( i+1\right) }=(G,\mathcal{E}^{\left( i\right) }G).$

Such a group is represented by the nilpotent matrices. Based on this
definition, it is possible to prove that every nilpotent group is solvable
[40, page 112]. These results lead us to the Lie-Kolchin theorem of major
importance

\ 

\textbf{Theorem 3.12.} (Lie and Kolchin, Ref.[49, p.113]) \textit{Let }$G$%
\textit{\ be connected solvable \ algebraic group acting on a projective
variety X. Then }$G$\textit{\ has a fixed point in }$X$\textit{.}

In view of the Remark 3.8., we know that such $G$ is a subgroup of $GL_{n}($K%
$).$ Moreover, $GL_{n}($K$)$ has at least another subgroup, called $%
semisimple$, for which Theorem 3.12. does not hold. In this case we have the
following

\textbf{Definition 3.13}. The group $G$ is \textit{semisimple} if it has no
closed connected commutative normal subgroups other than $e$.

Such a group is represented by the semisimple, i.e. diagonal (or torus),
matrices while the members of the unipotent group are represented by the
upper triangular matrices with all diagonal entries being equal to 1. In
view of the Theorem 3.12, the unipotent group is also solvable and,
accordingly, there must be an element \ $B$ of such a group fixing the flag $%
\ f_{0}$ defined by Eq.(3.11), i.e. $Bf_{0}=f_{0}.$ Let now $g\in GL_{n}($%
\textit{K}$)$. Then, naturally, $gf_{0}=f$ where $f\neq f_{0}.$ From here we
obtain, $f_{0}=g^{-1}f$. Next, we obtain as well, $Bg^{-1}f=g^{-1}f$ and,
finally, $gBg^{-1}f=f$. Based on these results, it follows that $gBg^{-1}=%
\tilde{B}$ is also an element \ of $GL_{n}($K$)$ fixing the flag $\ f$, etc.
This means that all such elements are conjugate to each other and form the $%
\emph{Borel}$ $\emph{subgroup}$. We shall denote all elements of this sort
by $B.$ These are made of upper triangular matrices belonging to $GL_{n}($K$%
).$ Surely, such matrices satisfy Proposition 3.9. The quotient group $G/B$
will act transitively on $X$. Since this quotient is also a linear algebraic
group, it is as well a projective variety called the \textit{flag variety,}
Ref.[48, p.176]

\textbf{Remark 3.14. }The flag variety is directly connected with the 
\textit{Schubert variety, }Ref\textit{.}[50, p.124].The Schubert varieties
were considered \ earlier, in our work, Ref.[51\textbf{]}, in connection
with the exact combinatorial solution of the Kontsevich-Witten (K-W) model.
Hence, the above remark naturally leads us to the combinatorial approach to
problems we are discussing in this part of our work and in Part IV.
Additional details on connections with K-W model will become apparent in
Section 4.

By now it should be clear that the group $G$ is made out of at least two
subgroups: $B,$ just described, and $N$. The \textit{maximal torus} $T$
subgroup of $G$ can be defined now as $T=B\cap N$. This fact allows\ us to
define the Weyl group $W=N/T$. Although this group has the same name as that
discussed in the Appendix, its true meaning in the present context requires
some explanations. They will be provided below.

This is done in several steps. First, using results of Appendix we notice
that the \textquotedblright true\textquotedblright\ Weyl group is made of
reflections, i.e. involutions of order 2. Following Tits, Ref.[32], we
introduce a quadruple $(G,B,N,S)$ \ (the Tits system) where $S$ is the
subgroup of $W$ \ made of elements such that $S=S^{-1}$ and $1$ $\notin S.$
Such a subroup always exists for the compact Lie groups considered as
symmetric spaces. Then, it can be shown that $G=BWB$ (\textit{Bruhat
decomposition}) \ and, moreover, that the Tits system is isomorphic to the
Coxeter system, i.e. to the Coxeter reflection group. The full proof can be
found in the monograph by Bourbaki, Ref.[32], Chr.6, paragraph 2.4.

Second, since $W=N/T$, it is of interest to see the connection (if any)
between $W$ and the quotient $G/B=BWB/B$ $=\left[ B\left( N/T\right) B\right]
/B.$ In view of the fact that $T=B\cap N,$ suppose that $N$ commutes with $B$%
. Then we would have $G/B\simeq \left( N/T\right) B$ and, since $B$ fixes
the flag $\ \ f$, we are left with the action of $N$ on the flag. In view of
the rule $M\circ f(\mathbf{z})=f(M\mathbf{z}),$ and noticing that the
diagonal matrix $T$ (the centralizer) can be chosen as a reference
(identity) transformation, we conclude that the commuting matrix $N$ (the
normalizer) should permute $t^{a_{i}}$. Consider an application of this rule
to the monomial $\mathbf{u}^{\mathbf{l}}=u_{1}^{l_{1}}\cdot \cdot \cdot
u_{n}^{l_{n}}\equiv z_{1}^{l_{1}a_{1}}\cdot \cdot \cdot z_{n}^{l_{n1}a_{n}}$%
. For such a map the character $c(t)$ is given by 
\begin{equation}
c(t)=t^{<\mathbf{l}\cdot \mathbf{a>}},  \tag{3.12}
\end{equation}%
where, in accord with Eq.(3.5), $<\mathbf{l}\cdot \mathbf{a}%
>=\sum\nolimits_{i}l_{i}a_{i}$ with both $l_{i}$ and $a_{i}$ being some
integers. Following Ref.[52], let us consider the limit $t\rightarrow 0$ in
the above expression. Clearly, we obtain: 
\begin{equation}
c(t)=\left\{ 
\begin{array}{c}
1\text{ if }<\mathbf{l}\cdot \mathbf{a}>=0 \\ 
0\text{ if }<\mathbf{l}\cdot \mathbf{a}>\neq 0%
\end{array}%
\right. .  \tag{3.13}
\end{equation}%
Evidently, the equation $<\mathbf{l}\cdot \mathbf{a}>=0$ describes a
hyperplane or, better, a set of hyperplanes for a given vector $\mathbf{a}$.
In view of Eq.(3.5), such a set forms at least one polyhedral cone (or
chamber in the terminology of Appendix). These results can be complicated a
little bit by introducing a subset $I$ $\subset $ $\{1,...,n\}$ such that,
say, only those $l_{i}^{\prime }s$ which belong to this subset satisfy $\ <%
\mathbf{l}\cdot \mathbf{a}>=0.$ Naturally, one obtains the one- to- one
correspondence between such subsets and earlier defined flags. Clearly, the
set of such constructed monomials forms an invariant of the torus group
action as discussed in in Part II. It remains to demonstrate that the Weyl
group $W=N/T$ permutes $a_{i}$'s thus forming an orbit transitively
\textquotedblright visiting\textquotedblright\ different hyperplanes. This
will be demonstrated momentarily. Before doing this, we would like to change
the rules of the game slightly\footnote{%
Such change of rules is consistent with arguments by Kirillov to be
discussed in the next subsection.}. To this purpose, we would like to
replace the limiting $t\rightarrow 0$ procedure by the procedure requiring $%
t\rightarrow \xi $ with $\xi $ being the nontrivial $n$-th root of unity.
After such a replacement we are entering the domain of the pseudo-reflection
groups discussed in Part II. Thus, replacing $t$ by $\xi $ causes us to
change the rule, Eq.(3.13), as follows: 
\begin{equation}
c(\xi )=\left\{ 
\begin{array}{c}
1\text{ if }<\mathbf{l}\cdot \mathbf{a}>=0\text{ }\func{mod}\text{\textit{n}}
\\ 
0\text{ if \ \ \ \ \ \ \ \ \ }<\mathbf{l}\cdot \mathbf{a}>\neq 0%
\end{array}%
\right. .  \tag{3.14}
\end{equation}%
At this point it is appropriate to recall Eq.(I,3.11a). In view of this
equation, we shall call the equation $<\mathbf{l}\cdot \mathbf{a}>=n$ \ as
the \textit{Veneziano condition} while the $\mathit{Kac-Moody-Bloch-Bragg}$ $%
\mathit{(K-M-B-B)}$ \textit{condition},$\mathit{\ }$Eq.(I, 3.22), can be
written now as $\ \ \mathit{<}\mathbf{l}\mathit{\cdot }\mathbf{a}\mathit{>=0}
$ $\func{mod}\mathit{n}$ \ .

The results of Appendix (part c)) indicate that the first option (the
Veneziano condition) is characteristic for the standard Weyl-Coxeter
(pseudo) reflection groups while the second is characteristic for the affine
Weyl-Coxeter groups thus leading to the Kac-Moody affine Lie algebras as
discussed in Part II.

At this moment we are ready to demonstrate that $W=N/T$ is indeed the Weyl
reflection group. Even though the full proof can be found, for example, in
the monograph by Bourbaki, Ref.[32], still it is instructive to provide
qualitative arguments exhibiting the essence of the proof (different from
that given by Bourbaki who use the Tits system).

Let us begin with an assembly of $\left( d+1\right) \times \left( d+1\right) 
$ matrices with complex coefficients.They belong to the group $GL_{d+1}(%
\mathbf{C}).$ Consider a subset of all diagonal matrices and, having in mind
physical applications, let us assume that the diagonal entries are made of $%
n-th$ roots of unity \ $\xi .$ Taking into account the results on
pseudo-reflection groups as discussed in Appendix (part d)) to Part II, each
diagonal entry can be represented by $\xi ^{k}$ with $1\leq k\leq n-1$ so
that there are $\left( n-1\right) ^{d+1}$ different diagonal matrices- all
commuting with each other. Among these commuting matrices we would like to
single out those which have all \ $\xi ^{k}$ $^{\prime }s$ the same.
Evidently, there are $n-1$ of them. They are effectively the unit matrices
and they are forming the centralizer of $W$. The rest belongs to the
normalizer.\footnote{%
As with Eq.(3.12), one can complicate matters by considering matrices which
have several diagonal entries which are the same. Then, as before, one
should consider the flag system where in each subsystem the entries are all
different. The arguments \ applied to such subsystems will proceed the same
way as in the main text.} The number $\left( n-1\right) ^{d+1}$ $/(n-1)=$ $%
\left( n-1\right) ^{d}$ was obtained earlier, e.g. see the discussion whoch
follows Eq.(1.7) ( and replace $2m$ by $n$) and the discussion which follows
this equation. This is not just a mere coincidence. In the next section we
shall provide some refinements of this result motivated by physical
considerations. It should be clear already that we are discussing only the
simplest possibility (of cubic symmetry) for the sake of illustration of
general principles. Clearly, the zonotope construction, introduced earlier
allows us to transfer our reasonings to more general cases.

Next, let us consider just one of the diagonal matrices \ $\tilde{T}$ whose
entries are all different and are made of powers of $\xi $. Let $g\in
GL_{d+1}(\mathbf{C})$ and consider an automorphism: $\mathcal{F}(\tilde{T}%
):=g\tilde{T}g^{-1}$. Along with it, we would like to consider an orbit $O(%
\tilde{T}):=g\tilde{T}C$ where $C$ is any of the diagonal matrices belonging
to earlier discussed centralizer.\footnote{%
Presence of $C$ factor underscores the fact that we are considering the
orbit of the factorgroup $W=N/T$.} Clearly, $O(\tilde{T})=g\tilde{T}g^{-1}gC=%
\mathcal{F}(\tilde{T})gC=\mathcal{F}(\tilde{T})C.$ Denote now $\tilde{T}$ =$%
\tilde{T}_{1}$ and consider another matrix $\tilde{T}_{2}$ belonging to the
same set and suppose that there is such matrix $g_{12}$ that $\tilde{T}_{2}C=%
\mathcal{F}(\tilde{T})C.$ If such a matrix exist, it should belong to the
normalizer and, naturally, the same arguments can be used to $\tilde{T}_{3}$%
, etc. Hence, the following conclusions can be drawn. If we had started with
some element \ $\tilde{T}_{1}$ of the maximal torus, the orbit of this
element will return back and intersect the maximal torus in \textit{finite}
number of points (in our case the number of points is exactly $\left(
n-1\right) ^{d}).$ By analogy with the theory of dynamical systems, we can
consider these intersection points of the orbit $O(\tilde{T})$ with the \ $T$%
-plane as the Poincare$^{\prime }$ crossections. Hence, as it is done in the
case of dynamical systems (e.g. see Section 2.2) we have to study the
transition map between these crossections. The orbit associated with such a
map is precisely the orbit of the Weyl group $W$. It acts on these points
transitively [49, p.147], as required. Provided that the set of fixed point
exists, such arguments justify the dynamical interpretation of the Weyl's
character formula presented in Section 2.2. The fact that such fixed point
set does exist is guaranteed by the Theorem 10.6. by Borel, Ref.[47]. Its
proof relies heavily on the Lie-Kolchin theorem\ (our Theorem 3.12.).

\subsection{Coadjoint orbits}

Thus far we were working with the Lie groups. To move forward we need to use
the Lie algebras associated with these groups. In what follows, we expect
our readers familiarity with basic relevant facts about the Lie groups \ and
Lie algebras which can be found in the books by Serre [53], Humphreys [54]
and Kac [36]. First, we notice that the Lie algebra matrices $h_{i}$
associated with the Lie group maximal tori $T_{i}$ (that is with all
diagonal matrices considered earlier) are commuting with each other thus
forming the Cartan subalgebra, i.e. 
\begin{equation}
\lbrack h_{i},h_{j}]=0.  \tag{3.15}
\end{equation}%
The matrices belonging to the normalizer are made of two types $x_{i}$ and $%
y_{i}$ corresponding to the root systems $\Delta ^{+}$ and $\Delta ^{-}$ \
defined in Appendix . The fixed point analysis described at the end of
previous section is translated into the following set of commutators 
\begin{align}
\left[ x_{i},y_{j}\right] & =\left\{ 
\begin{array}{c}
h_{i}\text{ if }i=j \\ 
0\text{ if }i\neq j%
\end{array}%
\right. ,  \tag{3.16a} \\
\left[ h_{i},x_{j}\right] & =<\alpha _{i}^{\vee },\alpha _{j}>x_{j}, 
\tag{3.16b} \\
\left[ h_{i},y_{j}\right] & =-<\alpha _{i}^{\vee },\alpha _{j}>y_{j}, 
\tag{3.16c}
\end{align}%
$i=1,...,n$. To insure that the matrices (operators) $x_{i}^{\prime }s$ and $%
y_{i}^{\prime }s$ are nilpotent (that is their Lie group ancestors belong to
the Borel subgroup $B$) one must impose two additional constraints.
According to Serre [53] these are: 
\begin{align}
\left( \text{ad}x_{i}\right) ^{-<\alpha _{i}^{\vee },\alpha _{j}>+1}(x_{j})&
=0,\text{ }i\neq j  \tag{3.16d} \\
\left( \text{ad}y_{i}\right) ^{-<\alpha _{i}^{\vee },\alpha _{j}>+1}(y_{j})&
=0,i\neq j.  \tag{3.16e}
\end{align}%
where ad$_{X}$ $Y=[X,Y].$\ \ \ From the book by Kac [36] one finds that 
\textit{exactly the same} relations characterize the Kac-Moody affine Lie
algebra. This fact is in accord with general results presented earlier in
this work and in Part II and is of major importance for develpment of our
formalism. In particular, for the purposes of this development it is
important to realize that for each $i$ Eq.s(3.16a-c) can be brought to form
(upon rescaling) coinciding with the Lie algebra $sl_{2}$(\textbf{C})%
\footnote{%
This fact is known as Jacobson-Morozov theorem [56].} and, if we replace 
\textbf{C} with any closed \ number field $\mathbf{F}$, then all semisimple
Lie algebras are made of copies of $sl_{2}$(\textbf{F}) [54, p.25].The Lie
algebra $sl_{2}$(\textbf{C}) is isomorphic to the algebra of operators
acting on differential forms living on the Hodge-type complex manifolds
[55]. This observation was absolutely essential for development of physical
applications in Part II.

Connections with Hodge theory can be also established through the method of
coadjoint orbits. We would like to discuss this method now. We begin by
considering an orbit in the Lie group. \ It is given by the Ad operator,
i.e. $O(X)=$Ad$_{g}X=gXg^{-1}$ where $g\in G$ and $X\in $\textsf{g} with $G$
being the Lie group and \textsf{g} its Lie algebra. For compact groups
globally and for noncompact locally every group element $g$ can be
represented via the exponential, e.g. $g(t)$=$\exp (tX_{g})$ with $X_{g}\in $%
\textsf{g.} Accordingly, for the orbit we can write $O(X)\equiv X(t)=\exp
(tX_{g})X\exp (-tX_{g}).$ Since the Lie group is a manifold $\mathcal{M}$,
the Lie algebra forms the tangent bundle of the vector fields at given point
of \ $\mathcal{M}$. In particular, the tangent vector to the orbit $X(t)$ is
determined, as usual, by $TO(X)=\frac{d}{dt}X(t)_{t=0}=[X_{g},X]=ad_{X_{g}}X.
$ Now we have to take into account that, actually, our orbit is made for a
vector $X$ \ coming from the torus, i.e. $T=\exp (tX).$ This means that when
we consider the commutator $[X_{g},X]$ it will be zero for $X_{g_{i}}=h_{i}$
and nonzero otherwise. Consider next the Killing form \ $\kappa (x,y)$\ for
two elements $x$ and $y$ of the Lie algebra: 
\begin{equation}
\kappa (x,y)=tr(adx\text{ }ady).  \tag{3.17}
\end{equation}%
From this definition it follows that 
\begin{equation}
\kappa ([x,y],z)=\kappa (x,[y,z]).  \tag{3.18}
\end{equation}%
The roots of the Weyl group can be rewritten in terms of the Killing form
[54]. Its purpose is to define the scalar multiplication between vectors
belonging to the Lie algebra and, as such, it allows one to determine the
notion of orthogonality between these vectors. In particular, if we choose $%
x\rightarrow X$ and $y,z$ $\in h_{i},$ then it is clear that the vector
tangential to the orbit $O(X)$ is going to be orthogonal to the subspace
spanned by the Cartan subalgebra. This result can be reinterpreted from the
point of view of symplectic geometry due to work of Kirillov [57].To this
purpose we would like to rewrite Eq.(3.18) in the equivalent form, i.e. 
\begin{equation}
\kappa (x,[y,z])=\kappa (x,\text{ad}_{y}z)=\kappa (\text{ad}_{x}^{\ast }y,z)
\tag{3.19}
\end{equation}%
where in the case of compact Lie group ad$_{x}^{\ast }y$ actually coincides
with ad$_{x}y$ . The reason for introducing the asterisk * lies in the
following chain of arguments. In Eq.(A.1) of Appendix (and in Eq.(3.5)) we
introduced vectors from the dual space. Such a construction is possible as
long as the scalar multiplication is defined. Hence, for the orbit Ad$_{g}X$
there must be a vector $\xi $ in the dual space \textsf{g}$^{\ast }$ such
that equation 
\begin{equation}
\kappa (\xi ,\text{Ad}_{g}X)=\kappa (\text{Ad}_{g}^{\ast }\xi ,X)  \tag{3.20}
\end{equation}%
defines the \textit{coadjoint orbit} $O^{\ast }$($\xi $)=Ad$_{g}^{\ast }\xi .
$ Accordingly, for such an orbit there is also the tangent vector $TO^{\ast }
$($\xi $)=$ad_{\mathsf{g}}^{\ast }\xi $ to the orbit and, clearly, we have $%
\kappa (\xi ,ad_{X_{g}}X)=\kappa (ad_{\mathsf{g}}^{\ast }\xi ,X).$ In the
case if we are dealing with the flag space, the family of coadjoint orbits
will represent the flag space structure as well. Next, let $x\in $\textsf{g}$%
^{\ast }$ and $\xi _{1},\xi _{2}\in TO^{\ast }$($x$), then consider the
properties of the (symplectic) form $\omega _{x}(\xi _{1},\xi _{2})$ to be
determined explicitly momentarily. For this purpose we need to introduce
some notations, e.g. $ad_{\mathsf{g}}^{\ast }x=f(x,\mathsf{g}),etc.$ so that
for \textsf{g}$_{1}$ and \textsf{g}$_{2}\in $\textsf{g} one has $\xi
_{i}=f(x,\mathsf{g}_{i}),$ $i=1,2.$ Then, one can claim that for the compact
Lie group and the associated with it Lie algebra $\omega _{x}(\xi _{1},\xi
_{2})=\kappa (x,[\mathsf{g}_{1},\mathsf{g}_{2}]).$ Indeed, using Eq.(3.18)
we obtain: $\kappa (x,[\mathsf{g}_{1},\mathsf{g}_{2}])=\kappa (\xi _{1},%
\mathsf{g}_{2})=-\kappa (x,[\mathsf{g}_{2},\mathsf{g}_{1}])=-\kappa (\xi
_{2},\mathsf{g}_{1}).$ Thus constructed form defines the symplectic
structure on the coadjoint orbit $O^{\ast }$($x$) since it is closed, skew
-symmetric, nondegenerate and is effectively independent of the choice of 
\textsf{g}$_{1}$ and \textsf{g}$_{2}.$The proofs can be found in the
literature [58].The obtained symplectic manifold $\mathcal{M}_{x}$ is the
quotient \textsf{g}$/\mathsf{g}_{h}$ with \textsf{g}$_{h}$ being made of
vectors of theCartan subalgebra so that for such vectors, by construction, $%
\omega _{x}(\xi _{1},\xi _{2})=0$. From the point of view of symplectic
geometry, the points for which $\omega _{x}(\xi _{1},\xi _{2})=0$ correspond
to the critical points for the velocity vector field on the manifold $%
\mathcal{M}_{x}$. I.e. these are the points at which the velocity field
vanishes. They are in one-to one correspondence with the fixed points of the
orbit $O(X).$ This fact allows us to use the Poincare$^{\prime }$-Hopf index
theorem (earlier used in our works on dynamics of 2+1 gravity [59]) in order
to obtain the Euler characteristic $\chi $ \ for such manifold as the sum of
indices of vector fields existing on $\mathcal{M}_{x}$.We shall provide more
details related to this observation below in Section 3.4.

To complete the above discussion, following Atiyah [60], we notice that
every nonsigular algebraic variety in projective space is symplectic. The
symplectic (K\"{a}hler) structure is inherited from that in the projective
space. The complex K\"{a}hler structure for the symplectic (Kirillov)
manifold is actually of the Hodge -type. This comes from the following
observations. First, since we have used the Killing form to determine the
Kirillov symplectic form $\omega _{x}$ and since the same Killing form is
used for the Weyl reflection groups [58], the induced unitary\ one
dimensional representation of the torus subgroup of $GL_{n}(\mathbf{C})$ is
obtained \ according to Kirillov [57] by simply replacing $t$ by the root of
unity in Eq.(3.12). This is permissible if and only if the integral of
two-form $\int\nolimits_{\gamma }\omega _{x}$ taken over any two dimensional
cycle $\gamma $ on the coadjoint orbit $O^{\ast }$($x$) is an integer. But
this is exactly the condition which makes the K\"{a}hler complex structure
that of the Hodge type [55].

\subsection{Construction of the moment map using methods of linear
programming}

In this subsection we are not employing the definition of the moment mapping
used in symplectic geometry [61]\footnote{%
Evidently, we are using the same thing anyway.}. Instead, we shall rely
heavily on works by Atiyah [60,62] with only slightest refinement coming
from noticed connections with the linear programming not mentioned in his
papers and in literature on symplectic geometry. In our opinion, such a
connection is helpful for better physical understanding of mathematical
methods discussed in this paper which might be useful for applications in
other disciplines.

Using Definition 1.1. of Section 1. we call the subset of $\mathbf{R}^{n}$ a
polyhedron \ $\mathcal{P}$ if there exist $m\times n$ matrix $\ A$ $($with $%
m<n)$ and a vector $\mathbf{b}\in \mathbf{R}^{m}$ such that according to
Eq.(1.8) we have 
\begin{equation*}
\mathcal{P}=\{\mathbf{x}\in \mathbf{R}^{n}\mid \mathbf{Ax}\leq \mathbf{b}\}.
\end{equation*}%
Since each component of the inequalities $\mathbf{Ax}\leq \mathbf{b}$
determines the half space while the equality $\mathbf{Ax}=b$ -the underlying
hyperplane, the polyhedron is an intersection of finitely many halfspaces.
The problem of linear programming can be formulated as follows [63] $:$ for
the linear functional $\mathcal{\tilde{H}[}\mathbf{x}]$=$\mathbf{c}^{\text{T}%
}\cdot \mathbf{x}$ find $\max $ $\mathcal{\tilde{H}[}\mathbf{x}]$ on $%
\mathcal{P}$ provided that the vector \textbf{c} is assigned. It should be
noted that this problem is just one of many related problems. It was
selected only because of its immediate relevance. Its relevance comes from
the fact that the extremum of $\mathcal{\tilde{H}[}\mathbf{x}]$ is achieved
at least at one of the vertices of $\mathcal{P}$. The proof of this we omit
since it can be found in any standard textbook on linear programming, e.g.
see [64] and references therein. This result does not require the polyhedron
to be centrally symmetric. Only convexity of the polyhedron is of
importance. This is physically plausible since, for instance, reflexive
polyhedra discussed in Section 1 in connection with mirror symmetry \ do not
require such central symmetry as can be seen from two dimensional examples
presented in Ref.[65\textbf{, }p.100].

To connect this optimization problem with results of our paper we constrain $%
\mathbf{x}$ variables to integers, i.e. to \textbf{Z}$^{n}.$ Such a
restriction is known in literature as \textit{integer linear programming}.
In our case, it is equivalent to considering symplectic manifolds of
Hodge-type (e.g. read page 11 of Atiyah's paper, Ref.[60] ).Hence, existence
of mirror symmetry as well as the method of coadjoint orbits both require
the underlying symplectic manifolds to be of Hodge-type. This has a deep
physical reason which will become clear when we shall discuss the
Khovanskii-Pukhlikov correspondence in the next section.

As a warm up exercise, following Fulton, Ref.[40], let us consider a
deformation retract of complex projective space \textbf{CP}$^{n}$ which is
the simplest possible toric variety [40,41]\footnote{%
Although such a construction was introduced in Part II, we write it down
explicitly anyway for the sake of uninterrupted reading.}. Such a retraction
is achieved by using the map : 
\begin{equation*}
\text{ }\tau :\text{\ }\mathbf{CP}^{n}\rightarrow \mathbf{P}_{\geq }^{n}=%
\mathbf{R}_{\geq }^{n+1}\setminus \{0\}/\mathbf{R}^{+}
\end{equation*}%
explicitly given by 
\begin{equation}
\tau :\text{ }(z_{0},...,z_{n})\mapsto \frac{1}{\sum\nolimits_{i}\left\vert
z_{i}\right\vert }(\left\vert z_{0}\right\vert ,...,\left\vert
z_{n}\right\vert )=(t_{0},...,t_{n})\text{ , }t_{i}\geq 0.  \tag{3.21}
\end{equation}%
The map $\tau $ by design is onto the standard $n$-simplex : $t_{i}\geq 0$, $%
t_{0}+...+t_{n}=1.$\ To bring physics to this discussion, let us consider
the Hamiltonian for the harmonic oscillator. In the appropriate system of
units we can write it as $\mathcal{H=}m(p^{2}+q^{2})$. More generally, for
finite set of oscillators, i.e. for the \textquotedblright
truncated\textquotedblright\ bosonic string, we have:$\ $\ $\mathcal{H}[%
\mathbf{z}]=\sum\nolimits_{i}m_{i}\left\vert z_{i}\right\vert ^{2},$ where,
following Atiyah [60], we introduced the complex $z_{j}$ variables via $%
z_{j}=p_{j}+iq_{j}.$ \ Let now such a Hamiltonian system (the truncated
string) possess the finite fixed energy $\mathcal{E}$. Then we obtain: 
\begin{equation}
\mathcal{H}[\mathbf{z}]=\sum\nolimits_{i=0}^{n}m_{i}\left\vert
z_{i}\right\vert ^{2}=\mathcal{E}\text{.}  \tag{3.22}
\end{equation}%
It is not difficult to realize that the above equation actually represents
the \textbf{CP}$^{n}$ since the points $z_{j}$ can be identified with the
points $e^{i\theta }z_{j}$ in Eq.(3.22) while keeping the above expression
form- invariant. In such a case one is saying that the \textit{reduced}
phase space for this model is \textbf{CP}$^{n}$ as discussed in Section 7.6.
of Part II.We can map such a model of \textbf{CP}$^{n}$ back into the
simplex using the map $\tau $. Since \textbf{CP}$^{n}$ is the simplest toric
variety [40,41], if we let $z_{j}$ to \textquotedblright
live\textquotedblright\ in such a variety it will be affected by the torus
action as discussed \ earlier in this section. This means that, in general,
the masses in Eq.(3.22) may change and, accordingly, the energy. Only if we
constrain the torus action to the simplex (or, more generally, to the
polyhedron as described by Fulton, Ref.[40],) will the energy be conserved.
Evidently, such a constraint is compatible with the original idea of
identification of points $z_{j}$ with $e^{i\theta }z_{j}$ . The fixed points
of \ such defined torus action\ are roots of unity according to Eq.(3.10).
In general, the existence of at least one fixed point is guaranteed for the
linear algebraic group by the Theorem 10.6, Ref.[47]. With such defined
torus action, $\left\vert z_{i}\right\vert ^{2}$ is just some \textit{%
positive} number, say, $x_{i}$. \textit{The essence of the moment map lies
exactly in such} \textit{identification\footnote{%
More accurate definition is given in Section 4.}}. Hence, we obtain the
following image of the moment map: 
\begin{equation}
\mathcal{\tilde{H}}[\mathbf{x}]=\sum\nolimits_{i=0}^{n}m_{i}x_{i}, 
\tag{3.23}
\end{equation}%
where we have removed the energy constraint for a moment thus making $%
\mathcal{\tilde{H}}[\mathbf{x}]$ to coincide with \ earlier defined linear
functional to be optimized. Now we have to find a convex polyhedron on which
such a functional is going to be optimized. Thanks to works by Atiyah
[60,62] and Guillemin and Sternberg [66\textbf{]}, this task is completed
already. Naturally, the vertices of such a polyhedron are the critical
points of the moment map. Then, the theorem of linear programing stated
earlier guarantees that $\mathcal{\tilde{H}}[\mathbf{x}]$ achieves its
maximum at \ least at some of its vertices. Delzant [67] had demonstrated
that this is the case without use of linear programming language.

It is helpful to illustrate the essence of above arguments by employing
simple but important example originally discussed by Frankel [68]. Consider
a two sphere $S^{2}$ of unit radius, i.e. $x^{2}+y^{2}+z^{2}=1,$ and
parametrize this sphere using coordinates $x=\sqrt{1-z^{2}}\cos \phi $, $y=%
\sqrt{1-z^{2}}\sin \phi $ , $z=z$. In section 4 we shall demonstrate that
the Hamiltonian \ for the free particle \textquotedblright
living\textquotedblright\ on such a sphere is given by $\mathcal{H}[z]%
\mathcal{=}m\left( 1-z\right) $ so that equations of motion produce the
circles of latitude. These circles become (critical) points of equilibria at
the north and south pole of the sphere, i.e. for $z=\pm 1.$ Evidently, these
are the fixed points of the torus action. Under such circumstances our
polyhedron is the segment [-1,1] and its vertices are located at $\pm 1$ (to
be compared with \ discussion in Section 1.). The image of the moment map $%
\mathcal{H}[x]\mathcal{=}m\left( 1-x\right) $ acquires its maximum at $x=1$
and the value $x=1$ corresponds to \textit{two} \ polyhedral vertices
located at 1 and -1 respectively. \ This doubling feature was noticed and
discussed in detail by Delzant [67] whose work contains all needed proofs.
These can be considered as elaborations on much earlier results by Frankel
[68]. The circles on the sphere are represening the torus action (e.g. read
the discussion following Eq.(3.22)) so that dimension of the circle is half
of that of the sphere. This happens to be a general trend : the dimension of
the Cartan subalgebra (more accurately, the normalizer of the maximal torus)
is half of the dimension of the symplectic manifold $\mathcal{M}$ [39,67%
\textbf{]}. Incidentally, in the next subsection we shall see that the
integral of \ the Kirillov-Kostant symplectic two- form $\omega _{x}$ over $%
S^{2}$ is equal to 2 so that the complex structure on the sphere is that of
the Hodge type as required. Also, the symplectic two- form $\omega _{x}=0$
at two critical points. Generalization of this example to the multiparticle
case will be discussed below and in Section 4.

The results discussed thus far although establish connection between the
singularities of symplectic manifolds and polyhedra do not allow us to
discuss the fine details distinguishing \ between different polyhedra.
Fortunately, this has been to a large degree accomplished in Refs.[58,69].
Such a task is equivalent to classification of all finite dimensional
exactly integrable systems in accord with the Lie\ groups and Lie algebras
associated with them..

\subsection{Calculation of the Euler characteristic}

Using results just presented we are ready to calculate the Euler
characteristic of the projective algebraic variety following ideas by Hopf
[45] and Hopf and Samelson [46]. \ To begin, we notice that in the case of
vector fields on $S^{2}$ discussed in the previous subsection there are two
fixed points. The Poincare$^{\prime }$-Hopf fixed point theorem (extensively
used in our earlier work on 2+1 gravity, Ref.[59]) tells us that $\chi $ is
the sum of indices of the vector (or line) fields foliating manifold $%
\mathcal{M}$. In our case, the index of each critical point is known to be 1
so that $\chi =2$ as required\footnote{%
Incidentally, if following Delzant, Ref.[67], we divide the number of
polyhedral vertices by factor of 2, then using Eq.(1.7) with $2m$ replaced
by 1 we shall reobtain the result $\chi =2.$ More formally, we can say that
the cardinality $\left\vert G\right\vert $=$\frac{1}{2}\dim \mathcal{M}$.
That this is indeed the case in general was provern by Delzant.}. In the
case of S$^{2}$ the Darboux coordinates can be chosen as $\phi $ and $z$
with $0\leq \phi <2\pi $ and $z$ $\in $ [-1,1]. The volume form $\Omega $ is 
$d\phi \wedge dz$ so that the phase space is effectively the product $%
R\times S^{1}.$ We would like to construct now a dynamical system whose
Darboux coordinates are \{ $t_{1},...,t_{n};\phi _{1},...,\phi _{n}\}.$ If
in the case of $S^{2}$ the $z$ coordinate varied in the segment [-1,1], now
we shall assume that the point $\mathbf{t}=$\{$t_{1},...,t_{k}\}$ can vary
inside some polytope $\mathcal{P\subset }$ \textbf{R}$^{k}$ including its
boundaries. For our purposes, in view of Eq.(3.22), it is sufficient to
consider only some simplex $\Delta _{k}$ living in \textbf{R}$^{k}.$ This
happens when all masses in Eq.(3.22) are the same so that using Eq.s(3.21)
and (3.22) we obtain equation for the simplex. In the case of $S^{2}$ we can
think of $z$ coordinate as deformation retract for $S^{2}.$ One can say that 
$S^{2}$ is the \textit{inflated symplectic manifold} corresponding to the
segment [-1,1], i.e. $S^{2}\sim R\times S^{1}$. Accordingly, we can say that 
\textbf{CP}$^{k}\sim \Delta _{k}\times S^{1}\times \cdot \cdot \cdot S^{1}.$
The Darboux coordinates $\{$ $t_{1},...,t_{k};\phi _{1},...,\phi
_{k}\}\rightarrow \{t_{1}^{\frac{1}{2}}e^{i\phi _{1}},...,t_{k}^{\frac{1}{2}%
}e^{i\phi _{k}},\sqrt{1-\tsum\nolimits_{i}t_{i}}\}\equiv
\{z_{1},...,z_{k},z_{k+1}\},$ provided that $t_{1}+...+t_{k+1}=1.$ These
results are in accord with Eq.(3.21). Such a choice of coordinates realizes 
\textbf{CP}$^{k}$ as the space of equivalence classes 
\begin{equation}
\left\vert z_{1}\right\vert ^{2}+\cdot \cdot \cdot +\left\vert
z_{k}\right\vert ^{2}+\left\vert z_{k+1}\right\vert ^{2}=1\text{ , }%
z_{i}\sim e^{i\phi _{i}}z_{i}\text{ , }i=1-k\text{.}  \tag{3.24}
\end{equation}%
of the points lying on the sphere $S^{2k+1}$ in \textbf{C}$^{k+1}$ (we used
this realization of \textbf{CP}$^{k}$ already in Part II). In accord with
previous subsection, it is the reduced phase space (the reduced symplectic
manifold $M_{red}$) for our dynamical system.

Following \ Section 1., it is of interest to consider the inflated simplex $%
n\Delta _{k}$ living on the lattice \textbf{Z}$^{k}$. Accordingly, we can
consider the associated with it the inflated symplectic manifold $\mathcal{M}
$. The indices of critical points of such a manifold produce its Euler
characteristic $\chi .$ Irrespective to locations of critical points on such
a manifold, the point \textbf{t} should have coordinates such that $%
t_{1}+...+t_{k}=n.$ If $\mathcal{P}$ is the rational polytope these
coordinates should be some integers. Accordingly, one has to count the
number of \ solutions to the equation $t_{1}+...+t_{k}=n$ in order to
determine the number $p(k,n)$ of such critical points. This number we know
already since it is given by Eq.(1.2). Accordingly, for physically
interesting case associated with our interpretation of the Veneziano
amplitudes we obtain, $p(k,n)=\chi .$ These rather simple arguments are
useful to compare with extremely sophisticated proofs of the same result
using methods of algebraic geometry, e.g. see Refs.[40-42]. These methods
are of importance however in case if one is interested in computation of
some observables as it is done earlier, for example, for the
Witten-Kontsevich model [51]. More on this will be said below and in Part
IV. Obtained results provide us with tools needed for symplectic treatment
of the Veneziano amplitudes and for restoration of generating function
associated with these amplitudes. This is accomplished in the next section.

\section{Exact solution of the Veneziano model :symplectic treatment}

\bigskip

\subsection{The Moment map, the Duistermaat-Heckman formula and the
Khovanskii-Pukhlikov correspondence}

\subsubsection{General remarks}

We have mentioned already number of times mathematical connections between
the Veneziano amplitudes (and the associated with them Veneziano partition
function) and dynamical systems. We would like to summarize these results
now.\ First, already in Part I we emphasized that the development in this
series of work is motivated in part by two major obseravations. These are :
a) the unsymmetrized Veneziano amplitude can be looked upon as the Laplace
transform of the partition function obtained by quantization of \textit{%
finite} set of harmonic oscillators as described in the work by Vergne [3],
b) the unsymmetrized Veneziano amplitude can be interpreted as one of the
periods associated with homology cycles on the variety of Fermat-type. These
observations are sufficient for development of both symplectic and
supersymmetric approaches leading to restoration of the underlying physical
model producing \ the Veneziano-like amplitudes. In Part II we strengthened
these observations by invoking theorems by Solomon and Ginzburg (Theorems
2.2. and 2.5. respectively). Also, in Part II using results by Shepard and
Todd and Serre we provided enough eveidence for the Veneziano partition
function to be supersymmetric.Using these results we obtained exact solution
for the Veneziano model, i.e. we have obtained the partition/generating
function for this model whose observables are unsymmetrized Veneziano
amplitudes. In this work we provided additional details directing us towards
alternative (symplectic) interpretation of this partition function. These
include: a) zeta function by Ruelle, b) method of coadjoint orbits and c)
the moment map. Connections between supersymmetric and symplectic
descriptions can be deduced using well written monograph by Berline, Getzler
and Vergne [70]. In view of this, to avoid excessive size of our paper, it
is sufficient to emphasize only things of immediate relevance. In
particular, we would like to discuss now the Duistermaat-Heckman formula.

\subsubsection{\protect\bigskip The Duistermaat-Heckman formula}

Although the description of the Duistermaat-Heckman \ (D-H) formula can be
found in many places, we would like to discuss it now in connection with
earlier obtained results. To this purpose, using \ Subsection 3.4. let us
consider once again the simplest dynamical model discussed there. The volume
form $\Omega $ for this model is given by $\Omega $ :=$d\theta \wedge dz$ so
that $\int\nolimits_{S^{2}}\Omega =4\pi $ as expected. With help of this
form the equation for the moment map can be obtained. According to the
standard rules [39,61\textbf{]}, given that $\xi =\frac{\partial }{\partial
\theta }$ , we obtain, 
\begin{equation}
i(\xi )\Omega =dz.  \tag{4.1}
\end{equation}%
The Hamiltonian $\mathcal{H}$, i.e. the moment map, is given in this case by 
$\mathcal{H}$=$z$. Consider now the integral $I(\beta )$ of the type 
\begin{equation}
I(\beta )=\int\limits_{\mathcal{M}_{red}}\tilde{\Omega}\exp (-\beta \mathcal{%
H})=\frac{1}{\beta }(\exp (\beta )-\exp (-\beta )).  \tag{4.2}
\end{equation}%
In this integral the reduced phase space is $\mathcal{M}_{red}=S^{2}/S^{1}$
so that $\tilde{\Omega}=dz$ and, as before, $\ z\in \lbrack -1,1].$ Eq.(4.2)
is essentially the D-H formula! We would like to explain this fact in some
detail. In view of the results of Subsection 3.3. we know that the moment
map $\mathcal{H}$ achieves its extrema at the vertices of $\mathcal{P}$.
Since in our case $\mathcal{P}$ is the segment $[-1,1]$, indeed, $\mathcal{H}
$ achieves its extrema at both 1 and -1 so that the r.h.s. of Eq.(4.2) is in
fact the sum over the vertices of $\mathcal{P}$ taken with the appropriate
weights. The D-H formula provides exactly the same answer. Indeed let $%
M\equiv M^{2n}$ be a compact symplectic manifold equipped with the momentum
map $\Phi :M\rightarrow \mathbf{R}$ and the (Liouville) volume form $%
dV=\left( \frac{1}{2\pi }\right) ^{n}\frac{1}{n!}$ $\Omega ^{n}.$\ According
to the Darboux theorem, the two-form $\Omega $ can be presented locally as: $%
\Omega =\sum\nolimits_{l=1}^{n}$\ \ $dq_{l}$\ $\wedge dp_{l}$\ . Suppose
that such a manifold has isolated fixed points $p$ belonging to the fixed
point set $\mathcal{V}$ associated with the isotropy subgroup $G$
(Definition 3.5.) acting on $M$. Then, in its most general form, the D-H
formula can be written as [39,61] 
\begin{equation}
\int\limits_{M}dVe^{\Phi }=\sum\limits_{p\in \mathcal{V}}\frac{e^{\Phi (p)}}{%
\prod\nolimits_{j}a_{j,p}}  \tag{4.3}
\end{equation}%
where $a_{1,p},...,a_{n,p}$ are the weights of the linearized action of $G$
on $T_{p}M$ . Using the Morse theory, Atiyah [62] and others [61] have
demonstrated that it is sufficient to keep terms up to quadratic in the
expansion of $\Phi $ around given $p$. In such a case the moment map looks
exactly like that given \ in Eq.(3.22). Moreover, the coefficients $%
a_{1,p},...,a_{n,p}$ are just \textquotedblright masses\textquotedblright\ $%
m_{i}$ in Eq.(3.22). We put quotation marks around masses since they can be
both positive and negative. With these remarks, it should be obvious that
Eq.(4.2) is the D-H formula. It should be noted that although in Eq.(4.3)
the space $M$ is not reduced, Eq.(4.2) can be written without requirement of
reduction as well. For this it is sufficient to consider in Eq.(4.2) the
form $\Omega $ =$\frac{1}{2\pi }d\theta \wedge dz.$ Hence, indeed, Eq.(4.2) 
\textit{is} the D-H formula. Consider now the limiting case $\beta
\rightarrow 0^{+}$ of Eq.(4.2). Then, we obtain 
\begin{equation}
I(\beta \rightarrow 0^{+})=2.  \tag{4.4}
\end{equation}%
But 2 is the \textit{Euclidean volume} of the polytope $\mathcal{P}$ , in
our case, the length of the segment $[-1,1].$ This is in accord with general
result obtained by Atiyah [60]. Now we would like to generalize this
apparently trivial result in several directions. First, we would like to
blow up the sphere so that its diameter would be $2m$. Second, we would like
to consider a collection of such spheres with respective diameters $2m_{i}$, 
$i=1-d$. For such a collection we can consider 2 situations : a) the total
energy $\mathcal{E}$ for the Hamiltonian $\mathcal{H}$=$\sum%
\nolimits_{i}z_{i}$ is not conserved \ and b) the total energy is conserved,
e.g. see Eq.(3.22). The first case is nonphysical but, apparently, is
relevant for theory of singularities of differentiable maps and is related
to the computation of the Milnor number. This issue was discussed in our
earlier work, Ref.[71]. The second case is physically relevant. Hence, we
would like to discuss it in some detail. Both cases can be found as
exercises on page 50 in the book by Guillemin, Ref.[38]. In discussing the
second case both Guillemin, Ref.[38], and Audin, Ref.[61 ], notice that the
action for the torus $T^{d}=S^{1}\times \cdot \cdot \cdot \times S^{1}$ on
such Hamiltonian system is diagonal (e.g. Section 3) and \ is made of $d$%
-tuples ($e^{i\theta _{1}},...,e^{i\theta _{d}})$ subject to the constraint $%
e^{i\theta _{1}}\cdot \cdot \cdot e^{i\theta _{d}}=1.$ This constraint is
actually the Veneziano condition discussed in Section 3, e.g. see Eq.(3.14).

Based on the information just mentioned, we would like to be more specific
now. To this purpose, following Vergne [72] and Brion [6] we would like to
consider the simplest nontrivial case of the integral of the form 
\begin{equation}
I(k)=\int\nolimits_{k\Delta }dx_{1}dx_{2}\exp \{-(y_{1}x_{1}+y_{2}x_{2})\} 
\tag{4.5}
\end{equation}%
where $k\Delta $ is dilated standard simplex with coefficient of dilation $k$%
. Following these authors, calculation of \ this integral can be done
exactly with the result, 
\begin{equation}
I(k)=\frac{1}{y_{1}y_{2}}+\frac{e^{-ky_{1}}}{y_{1}(y_{1}-y_{2})}+\frac{%
e^{-ky_{2}}}{y_{2}(y_{2}-y_{1})}.  \tag{4.6}
\end{equation}%
As in earlier case of Eq.(4.2), the obtained result fits the D-H formula,
Eq.(4.3), and, as before, in the limit: $y_{1},y_{2}\rightarrow 0$ some
calculation produces the anticipated result $:Volk\Delta =k^{2}/2!$ , in
accord with Eq.(4.4). In view of results of Parts I, II and this work, this
integral is of relevance to calculation of Veneziano amplitudes (and it does
have symplectic meanning !): The standard simplex $\Delta $ in the present
case is just the deformation retract for the Fermat (hyper)surface
associated with calculation of the Veneziano (or Veneziano-like)
amplitudes.\ The relevance of this integral to the Veneziano amplitude is
far from superficial as we would like to discuss now.

\QTP{Body Math}
\bigskip

\subsubsection{The Khovanskii-Pukhlikov correspondence and calculation of $%
\protect\chi $}

The Khovanskii-Pukhlikov correspondence can be understood based on the
following generic example. Following Ref.[73] we would like to compare the
integral 
\begin{equation}
I(z)=\int\limits_{s}^{t}dxe^{zx}=\frac{e^{tz}}{z}-\frac{e^{sz}}{z}\text{ \
with the sum }S(z)=\text{ }\sum\limits_{k=s}^{t}e^{kz}=\frac{e^{tz}}{1-e^{-z}%
}+\frac{e^{sz}}{1-e^{z}},  \tag{4.7}
\end{equation}%
where Eq.(1.4) was used for calculation of S(z).One can pose a problem : is
there way to transform the integral $I$ into the sum $S$ ? Clearly, we are
interested in such a transform in view of the fact that the Ehrhart
polynomial computes the number of lattice points of the dilated polytope
while the D-H integral can be used only for calculation of the \textit{%
Euclidean }volumes of such polytopes as we just demonstrated on simple
examples. The positive answer to the above question was found by Khovanskii
and Pukhlikov [74] and refined by many others, e.g. see Ref.[73]. Before
discussing their work, we would like to write down the discrete analog of
the result, Eq.(4.6). It is given by 
\begin{align}
S(k)& =\frac{1}{1-e^{-y_{1}}}\frac{1}{1-e^{-y_{2}}}+\frac{1}{1-e^{y_{1}}}%
\frac{e^{-ky_{1}}}{1-e^{y_{1}-y_{2}}}+\frac{1}{1-e^{y_{2}}}\frac{e^{-ky_{2}}%
}{1-e^{y_{2}-y_{1}}}  \tag{4.8} \\
& =\sum\limits_{\left( l_{1},l_{2}\right) \in k\Delta }\exp
\{-(y_{1}l_{1}+y_{2}l_{2})\}.  \notag
\end{align}%
This result can be obtained rather straightforwardly using Brion's formula
for the generating function for polytopes. It is given by earlier discussed
Eq.(1.11) and, hence, it is in complete accord with this more general
equation. Some computational details can be found in the monograph by
Barvinok, Ref.[7]. \ Following his exposition we would like to discuss some
physics behind these formal calculations. For this we need to use the
definition of the monoid $S_{\sigma }$, Eq.(3.3), introduced earlier. In
view of the Remark 9.9.(Part II) the set $a_{1},...,a_{k}$ forms a basis of
the vector space $V$ \ so that the monoid $S_{\sigma }$ defines a \textit{%
rational} polyhedral cone $\sigma $. Thanks to the theorem by Brion [6,7]
the generating function in the l.h.s. of Eq.(1.11) can be conveniently
rewritten as 
\begin{equation}
f(\mathcal{P},\mathbf{x})=\sum\limits_{\mathbf{m}\in \mathcal{P}\cap \mathbf{%
Z}^{d}}\mathbf{x}^{\mathbf{m}}=\sum\limits_{\sigma \in Vert\mathcal{P}}%
\mathbf{x}^{\mathbf{\sigma }}  \tag{4.9a}
\end{equation}%
so that for the \textit{dilated} polytope it reads as follows 
\begin{equation}
f(k\mathcal{P},\mathbf{x})=\sum\limits_{\mathbf{m}\in k\mathcal{P}\cap 
\mathbf{Z}^{d}}\mathbf{x}^{\mathbf{m}}=\sum\limits_{i=1}^{n}\mathbf{x}^{%
\mathbf{kv}_{i}}\sum\limits_{\sigma _{i}}\mathbf{x}^{\mathbf{\sigma }_{i}}. 
\tag{4.9b}
\end{equation}%
In the last formula the summation is taking place over all vertices whose
location is given by the vectors from the set \{$\mathbf{v}_{1},...,\mathbf{v%
}_{n}\}.$ This means that in actual calculations one can first calculate the
contributions coming from the cones $\sigma _{i}$ of the undilated
(original) polytope $\mathcal{P}$ and only then one can use this equation in
order to get the result for the dilated polytope. Let us apply these general
rules to our specific problem of computation of $S(k)$ in Eq.(4.8). We have
our simplex with vertices in x-y plane given by the vector set \{ \textbf{v}$%
_{1}$=$(0,0)$, \textbf{v}$_{2}$=(1,0), \textbf{v}$_{3}$=(0,1)\} where we
have written the x coordinate first. For this case we have 3 cones: $\sigma
_{1}=l_{2}\mathbf{v}_{2}+l_{3}\mathbf{v}_{3}$ ; $\sigma _{2}=\mathbf{v}%
_{2}+l_{1}(-$\textbf{v}$_{2})+l_{2}($\textbf{v}$_{3}-$\textbf{v}$%
_{2});\sigma _{3}=$\textbf{v}$_{3}+l_{3}($\textbf{v}$_{2}-$\textbf{v}$%
_{3})+l_{1}(-\mathbf{v}_{3});$\{$l_{1}$, $l_{2}$ , $l_{3}$ \}$\in \mathbf{Z}%
_{+}$ . In writing these expressions for the cones we have taken into
account that, according to Brion, when making calculations the apex of each
cone should be chosen as the origin of the coordinate system. Calculation of
contributions to generating function coming from $\sigma _{1}$ is the most
straightforward. Indeed, in this case we have $\mathbf{x}%
=x_{1}x_{2}=e^{-y_{1}}e^{-y_{2}}.$ Now, the symbol $\mathbf{x}^{\mathbf{%
\sigma }}$ in Eq.s(4.9) should be understood as follows. Since $\sigma _{i}$
, $i=1-3$, is actually a vector, it has components, like those for \textbf{v}%
$_{1},$etc$.$ We shall write therefore $\mathbf{x}^{\mathbf{\sigma }%
}=x_{1}^{\sigma (1)}\cdot \cdot \cdot x_{d}^{\sigma (d)}$ where $\sigma (i)$
is the i-th component of such a vector. Under these conditions calculation
of the contributions from the first cone with the apex located at (0,0) is
completely straightforward 
\begin{equation}
\sum\limits_{\left( l_{2},l_{3}\right) \in
Z_{+}^{2}}x_{1}^{l_{2}}x_{2}^{l_{3}}=\frac{1}{1-e_{{}}^{-y_{1}}}\frac{1}{%
1-e^{-y_{2}}}  \tag{4.10}
\end{equation}%
since it is reduced to the computation of the infinite geometric
progressions. But physically, the above result can be looked upon as a
product of two partition functions for two harmonic oscillators whose ground
state energy was discarded. By doing the rest of calculations in the way
just described we reobtain $S(k)$ from Eq.(4.8) as required. This time,
however, we know that the obtained result is associated with the assembly of
harmonic oscillators of frequencies $\pm y_{1}$, $\pm y_{2}$ and $\pm
(y_{1}-y_{2})$ whose ground state energy is properly adjusted. The
\textquotedblright frequencies\textquotedblright\ (or masses) of these
oscillators are coming from the Morse-theoretic considerations for the
moment maps associated with the critical points of symplectic manifolds as
explained \ in the paper by Atiyah \textbf{[}62\textbf{]}. These masses
enter into the \textquotedblright \textit{classical}\textquotedblright\ D-H
formula. It is just a classical partition function for a system of such
described harmonic oscillators living in phase space containing
singularities. The D-H classical partition function, Eq.(4.6), has its
quantum analog, Eq.(4.8), just described. The ground state for such a
quantum system is degenerate with degeneracy being described by the Kostant
multiplicity formula. To calculate this degeneracy would require us to study
the limiting case: $y_{1},y_{2}\rightarrow 0$ of Eq.(4.8) for $S(k).$
Surprisingly, unlike the continuum case sudied in the previous subsection,
calculation of number of points belonging to the dilated simplex $k\Delta $
(which is equivalent to the calculation of the Kostant multiplicity formula
or,which is the same, to the computation of the Ehrhart polynomial or to the
Euler characteristic $\chi $ of the associated projective toric variety) is
very nontrivial in the present case. It is facilitated by the observation
that in the limit $s\rightarrow 0$ the following expansion can be used 
\begin{equation}
\frac{1}{1-e^{-s}}=\frac{1}{s}+\frac{1}{2}+\frac{s}{12}+O(s^{2}).  \tag{4.11}
\end{equation}%
Rather lengthy calculation involving this expansion produces in the end the
anticipated result for the Ehrhart polynomial: 
\begin{equation}
\left\vert k\Delta \cap \mathbf{Z}^{2}\right\vert =P(k,2)=\frac{k^{2}}{2}+%
\frac{3}{2}k+1.  \tag{4.12}
\end{equation}%
Obtained results and their interpretations are in formal accord with those
by Vergne, Ref.[3]. In her work no details (like those presented above) or
physical applications are given however. At the same time, the results
obtained thus far apparently are not in agreement with earlier obtained
major result, Eq.(1.1). Fortunately, the situation can be corrected with
help of the following theorem by Barvinok, Ref.[75].

\bigskip

\textbf{Theorem 4.1. }\ \textit{For the fixed lattice of dimensionality} $d$ 
\textit{there exist a polynomial time algorithms which, for any given
rational polytope }$\mathcal{P}$, $\mathit{calculate}$ $\mathit{the}$ $%
\mathit{generating}$ $\mathit{function}$ $f(\mathcal{P},x)$ with the result 
\begin{equation}
f(\mathcal{P},x)=\sum\limits_{\mathbf{m}\in \mathcal{P}\cap \mathbf{Z}^{d}}%
\mathbf{x}^{\mathbf{m}}=\sum\limits_{i\in I}\epsilon _{i}\frac{\mathbf{x}%
^{p_{i}}}{\left( 1-\mathbf{x}^{a_{i\text{ }1}}\right) \cdot \cdot \cdot
\left( 1-\mathbf{x}^{a_{i\text{ }d}}\right) }  \tag{4.13}
\end{equation}%
\textit{where} $\epsilon \in \{-1,1\},p_{i},a_{i\text{ }j}\in \mathbf{Z}^{d}$
\textit{and} $a_{i\text{ }j}\neq 0$ $\forall $ $i,j$ . \textit{In fact,} $%
\forall $ $i$ , \textit{the set} $a_{i\text{ }1}$ $,...,$ $a_{i\text{ }d}$ 
\textit{forms a basis of} \textbf{Z}$^{d}$ \textit{and }$I$ \textit{is the
set }\{$1,...,n$\}\textit{labeling the} \textit{vertices of} $\mathcal{P}$.

\bigskip

\textbf{Remark 4.2. }It is easy to check this result using Eq.(1.1) for $n$\
(or $d$) equal to 2 and comparing it with $S(k)$ from Eq.(4.8). It should be
noted however, that Eq.(1.1) was obtained using some kind of combinatorial
and supersymmetric arguments as explained in Part II while Eq.(4.8) is
obtained exclusively based on use of the bosonic formalism. It should be
clear that both approaches leading to the design of new model reproducing
the Veneziano and Veneziano-like amplitudes can be used in principle since
they are essentially equivalent in view of the earlier mentioned Ref.[70].

At this point, finally, we are ready do discuss the Khovanskii-Pukhlikov
correspondence. It should be considered as alternative to the \ method of
coadjoint orbits discussed in Section 3.2. Naturally, both methods are in
agreement with each other with respect to final results. Following
Refs[38,72,73] we introduce the Todd operator (transform) via 
\begin{equation}
Td(z)=\frac{z}{1-e^{-z}}.  \tag{4.14}
\end{equation}%
In view of Eq.(4.7), it can be demonstrated [73] that 
\begin{equation}
Td(\frac{\partial }{\partial h_{1}})Td(\frac{\partial }{\partial h_{2}}%
)(\int\nolimits_{s-h_{1}}^{t+h_{2}}e^{zx}dx)\mid
_{h_{1}=h_{1}=0}=\sum\limits_{k=s}^{t}e^{kz}.  \tag{4.15}
\end{equation}%
This result can be now broadly generalized following ideas of Khovanskii and
Pukhlikov, Ref. [\textbf{66?}]. In particular, the relation 
\begin{equation}
Td(\frac{\partial }{\partial \mathbf{z}})\exp \left(
\sum\limits_{i=1}^{n}p_{i}z_{i}\right) =Td(p_{1},...,p_{n})\exp \left(
\sum\limits_{i=1}^{n}p_{i}z_{i}\right)   \tag{4.16}
\end{equation}%
happens to be the most useful. Applying it to 
\begin{equation}
i(x_{1},...,x_{k};\xi _{1},...,\xi _{k})=\frac{1}{\xi _{1}...\xi _{k}}\exp
(\sum\limits_{i=1}^{k}x_{i}\xi _{i})  \tag{4.17}
\end{equation}%
we obtain, 
\begin{equation}
s(x_{1},...,x_{k};\xi _{1},...,\xi _{k})=\frac{1}{\prod\limits_{i=1}^{k}(1-%
\exp (-\xi _{i}))}\exp (\sum\limits_{i=1}^{k}x_{i}\xi _{i}).  \tag{4.18}
\end{equation}%
This result should be compared now with the individual terms on the r.h.s.
of Eq.(1.11) on one hand and with the individual terms on the r.h.s of
Eq.(4.3) on another. Evidently, with help of the Todd transform the exact
\textquotedblright classical\textquotedblright\ results for the D-H integral
are transformed into the \textquotedblright quantum\textquotedblright\ Weyl
character formula.

We would like to illustrate these general observations by comparing the D-H
result, Eq.(4.6), with the Weyl character formula (e.g.see Eq.(1.11)),
Eq.(4.8). To this purpose we need to use already known data for the cones $%
\sigma _{i}$ , $i=1-3,$ and the convention for the symbol $\mathbf{x}%
^{\sigma }$. In particular, \ for the first cone we have already $:\mathbf{x}%
^{\sigma _{1}}=x_{1}^{l_{1}}x_{2}^{l_{2}}=\left[ \exp (l_{1}y_{1})\right]
\cdot \left[ \exp (l_{2}y_{2})\right] \footnote{%
To obtain correct results we needed to change\ signs in front of $l_{1}$ and 
$l_{2}$ . The same should be done for other cones as well.}.$ Now we
assemble the contribution from the first vertex using Eq.(4.6). We obtain, $%
\left[ \exp (l_{1}y_{1})\right] \cdot \left[ \exp (l_{2}y_{2})\right]
/y_{1}y_{2}.$ Using the Todd transform we obtain as well, 
\begin{equation}
Td(\frac{\partial }{\partial l_{1}})Td(\frac{\partial }{\partial l_{2}})%
\frac{1}{y_{1}y_{2}}\left[ \exp (l_{1}y_{1})\right] \cdot \left[ \exp
(l_{2}y_{2})\right] \mid _{l_{1}=l_{2}=0}=\frac{1}{1-e^{-y_{1}}}\frac{1}{%
1-e^{-y_{2}}}.  \tag{4.19}
\end{equation}%
Analogously, for the second cone we obtain: \textbf{x}$_{2}^{\sigma
}=e^{-ky_{1}}e^{-l_{1}y_{1}}e^{-l_{2}(y_{1}-y_{2})}$ \ so that use of the
Todd transform produces 
\begin{equation}
Td(\frac{\partial }{\partial l_{1}})Td(\frac{\partial }{\partial l_{2}})%
\frac{1}{y_{1}\left( y_{1}-y_{2}\right) }%
e^{-ky_{1}}e^{-l_{1}y_{1}}e^{-l_{2}(y_{1}-y_{2})}\mid _{l_{1}=l_{2}=0}=\frac{%
1}{1-e^{y_{1}}}\frac{e^{-ky_{1}}}{1-e^{y_{1}-y_{2}}},  \tag{4.20}
\end{equation}%
etc.

In Section 3.4. we sketched ideas behind calculations of Euler
characteristic $\chi $. It is instructive in the light of just obtained
results to reobtain $\chi .$

To accomplish the task is actually not difficult since it is based on the
information we have presented already. Indeed, by looking at the last two
equations it makes sense to rewrite formally the partition function,
Eq.(4.5), in the following symbolic form 
\begin{equation}
I(k,\mathbf{f})=\int\limits_{k\Delta }d\mathbf{x}\exp \mathbf{(-f\cdot x)} 
\tag{4.21}
\end{equation}%
valid for any finite dimension $d$. Since we have performed all calculations
explicitly for two dimensional case, for the sake of space, we only provide
the idea behind such type of calculation for any $d$ \footnote{%
Mathematically inclined reader is encoraged to read paper by Brion and
Vergne, Ref.[76], where all missing details are scrupulously presented.}. In
particular, using Eq.(4.3) we can rewrite this integral formally as follows 
\begin{equation}
\int\limits_{k\Delta }d\mathbf{x}\exp \mathbf{(-f\cdot x)=}\sum\limits_{p}%
\frac{\exp (-\mathbf{f}\cdot \mathbf{x(}p\mathbf{)})}{\prod%
\limits_{i}^{d}h_{i}^{p}(\mathbf{f})}.  \tag{4.22}
\end{equation}%
Applying the Todd operator (transform) to both sides of this formal
expression and taking into account Eq.s(4.19), (4.20) (providing assurance
that such an operation indeed is legitimate and makes sense) we obtain, 
\begin{align}
\int\limits_{k\Delta }d\mathbf{x}\prod\limits_{i=1}^{d}\frac{x_{i}}{1-\exp
(-x_{i})}\exp \mathbf{(-f\cdot x)}& =\sum\limits_{\mathbf{v}\in Vert\mathcal{%
P}}\exp \{<\mathbf{f}\cdot \mathbf{v}>\}\left[ \prod\limits_{i=1}^{d}(1-\exp
\{-h_{i}^{v}(\mathbf{f})u_{i}^{v}\})\right] ^{-1}  \notag \\
& =\sum\limits_{\mathbf{x}\in \mathcal{P\cap }\mathbf{Z}^{d}}\exp \{<\mathbf{%
f}\cdot \mathbf{x}>\}  \tag{4.23}
\end{align}%
where the last equation was written in view of Eq.(1.11). From here, it is
clear that in the limit : $\mathbf{f}=0$ we reobtain back $\chi $.

\subsection{\protect\bigskip From Riemann-Roch-Hirzebruch to Witten and
Lefschetz via Atiyah and Bott}

As it was noticed already by Khovanskii and Puklikov [74] and elaborated by
others, e.g. see Ref.[76], in the limit : $\mathbf{f}=0$ the integral in the
l.h.s of Eq.(4.21) can be associated with the Hirzebruch--Grotendieck-
Riemann- Roch formula for the Euler characteristic $\chi (E).$ In standard
notations [76,77] it is given by 
\begin{equation}
\chi (E)=\int\limits_{X}ch(E)\cdot Td(TX),  \tag{4.24}
\end{equation}%
where $E$ is a vector bundle over the variety $X$, $ch(E)$ is the Chern
character of $E$ and $Td(TX)$ is the Todd class of the tangent bundle $TX$ \
of \ $X$. This formula is too formal to be used immediately. The
mathematical formalism \ of equivariant cohomology is needed for actual
calculations connecting Eq.(4.22) with the l.h.s of Eq.(4.21). It was
developed in the classical paper by Atiyah and Bott, Ref.[25] inspired by
earlier work by Witten [26] on supersymmetry and Morse theory. In this work
we shall use only a small portion of their results. A very pedagogical
exposition of the results by Atiyah and Bott can be found in the monograph
by Guillemin and Sternberg [78] also containing helpful additional
supersymmetric information.

We begin our discussion with the following observations. Earlier, in Section
3.2 we introduced the Kirillov-Kostant symplectic two-form $\omega _{x}.$ We
noticed that this form is defined everywhere outside the set of critical
points of symplectic manifold $\mathcal{M}$. The simplest example of the
symplectic two- form was given in Section 3.4 for the case of two -sphere $%
S^{2}$ where it coincides with the volume form $\Omega =d\phi \wedge dz$ for
which $\int\nolimits_{S^{2}}\Omega =4\pi .$ At the same time, the \textit{%
symplectic} volume form is given by $\Omega /2\pi $ so that the integral
over $S^{2}$ becomes equal to 2. This fact reminds us about the Gauss-Bonnet
theorem for the 2-sphere which is prompting us to associate the two -form $%
\Omega /2\pi $ with the curvature two- form. To make things more interesting
we recall some facts from the differential analysis on complex manifolds as
described, for example, in the book by Wells, Ref.[55]. From this reference
we find that the first Chern class $c_{1}(E)$ of the $E$ vector bundle over
the sphere $S^{2}$ is given by 
\begin{equation}
c_{1}(E)=\frac{i}{\pi }\frac{dz\wedge d\bar{z}}{\left( 1+\left\vert
z\right\vert ^{2}\right) ^{2}}=\frac{2}{\pi }\frac{\rho d\rho d\phi }{\left(
1+\rho ^{2}\right) ^{2}}  \tag{4.25}
\end{equation}%
so that $\int\nolimits_{S^{2}}c_{1}(E)=2.$ Next, let us recall that \textit{%
any} K\"{a}hler manifold is symplectic [61] and that for any K\"{a}hler
manifold the second fundamental form $\Omega $ can be written locally as $%
\Omega =\frac{i}{2}\sum\nolimits_{ij}h_{ij}(z)dz_{i}\wedge d\bar{z}_{j}$ so
that $h_{ij}(z)=\delta _{ij}+O(\left\vert z\right\vert ^{2}).$ Hence, any
symplectic volume form can be rewritten in terms of just described form $%
\Omega .$ The form $\Omega $ is closed but not exact. Evidently, up to a
constant, $c_{1}(E)$ in Eq.(5.2) coincides with the standard K\"{a}hler two
-form. In view of the Gauss-Bonnet theorem, it is not exact. An easy
calculation shows that the form $d\phi \wedge dz$ can also be brought to the
standard K\"{a}hler two- form (again up to a constant). \ Moreover, for the
Hamiltonian of planar harmonic oscillator discussed in Section 3.3. the
standard symplectic two-form $\Omega $ can be written in several equivalent
ways 
\begin{equation}
\Omega =dx\wedge dy=rdr\wedge d\theta =\frac{1}{2}dr^{2}\wedge d\theta =%
\frac{i}{2}dz\wedge d\bar{z}  \tag{4.26}
\end{equation}%
and is certainly K\"{a}hlerian. For collection of $k$ harmonic oscillators
the symplectic two-form $\Omega $ is given, as usual, by $\Omega $ =$%
\sum\nolimits_{i=1}^{k}dx_{i}\wedge dy_{i}=\frac{i}{2}\sum%
\nolimits_{i=1}^{k}dz_{i}\wedge d\bar{z}_{i}$ so that its $n$-th power is
given by $\Omega ^{n}=\Omega \wedge \Omega \wedge \cdot \cdot \cdot \wedge
\Omega $ =$dx_{1}\wedge dy_{1}\wedge \cdot \cdot \cdot dx_{n}\wedge dy_{n}$.
In view of these results, it is convenient to introduce the differential
form 
\begin{equation}
\exp \Omega =1+\Omega +\frac{1}{2!}\Omega \wedge \Omega +\frac{1}{3!}\Omega
\wedge \Omega \wedge \Omega +\cdot \cdot \cdot \text{ \ \ \ .}  \tag{4.27}
\end{equation}%
By design, this expansion will have only $k$ terms. \ Our earlier discussion
of the moment map in Section 3.3. \ suggests that just described case of the
collection of harmonic oscillators is generic since its existence is
guaranteed by the Morse theory as discussed by Atiyah [62].\footnote{%
In view of earlier discussed examples, we are interested only in the
rotationally invariant observables, this means that the $\theta $ (or $\phi )
$ dependence \ in the two-form, Eq.(5.3), can be dropped \ which is
equivalent to considering only the reduced phase space.This is \ meanigful
both mathematically and physically. Details can be found in Ref.[33], pages
65-71.
\par
\bigskip } In view of Eq.(4.23) such an expansion can be formally associated
with the total Chern class. Hence, we shall associate $\exp \Omega $ with
the total Chern class. Since all symplectic manifolds we considered earlier
possess singularities the standard homology and cohomology theories should
be replaced by equivariant ones as explained by Atiyah and Bott, Ref.[25].
To this purpose we observe that in the absence of singularities the
symplectic 2-form $\Omega $ is always closed, i.e. $d\Omega =0.$ In case of
singularities, one should replace the exterior derivative $d$ by $\tilde{d}%
=d+i(\xi )\footnote{%
E.g. see Eq.(4.1).}$ while changing $\Omega $ to $\Omega -\mathbf{f\cdot x}$
in notations of of Eq.(4.21). The D-H integral, Eq.(4.21) can be formally
rewritten now as 
\begin{equation}
\int\limits_{k\Delta }d\mathbf{x}\exp \mathbf{(-f\cdot x)=}%
\int\limits_{k\Delta }\exp (\tilde{\Omega})  \tag{4.28}
\end{equation}%
where $\tilde{\Omega}=\Omega -\mathbf{f\cdot x.}$ The form $\tilde{\Omega}$
is eqivariantly closed. Indeed, since $\tilde{d}\tilde{\Omega}=d\Omega
+i(\xi )\Omega -\mathbf{f\cdot dx}$ then, in view of Eq.(4.1), $i(\xi
)\Omega -\mathbf{f\cdot dx}=0$ by design, while $d\Omega =0$ everywhere,
except at singularities (critical points) where $\Omega =0$ $($as discussed
in Section 3.2.$)$. Hence, $\tilde{d}\tilde{\Omega}=0$ as required. Since $%
\Omega $ can be identified with the Chern class one should identify $\mathbf{%
f\cdot x}$ with the Chern class as well, i.e. $\mathbf{f\cdot x\equiv }%
\sum\nolimits_{i=1}^{d}f_{i}c_{i}$ \ where we took into account that \ $%
E=L_{1}\oplus \cdot \cdot \cdot L_{d}$ \ because \textbf{C}$^{d}=\mathbf{C}%
\oplus \mathbf{C}\oplus \cdot \cdot \cdot \oplus $ $\mathbf{C}$ so that $%
L_{i}$ is the line bundle associated with $\mathbf{C}_{i}$ . After such an
identification Eq.(4.24) can be rewritten as 
\begin{equation}
\chi (E)=\int\limits_{X}e^{\Omega }\cdot \prod\limits_{i=1}^{d}\frac{c_{i}}{%
1-\exp (-c_{i})}.  \tag{4.29}
\end{equation}%
Obtained result is in agreement with that given in the book by Guillemin,
Ref.[38, p.60].

In view of the results of Part I, and Theorems 2.2. (by Solomon) and 2.5.
(by Ginzburg) of Part II one can achieve more by discussing the intersection
cohomology ring of the reduced spaces associated with the D-H measures.
Since in Part I we noticed already that the Veneziano amplitudes can be
formally associated with the period integrals for the Fermat (hyper)surfaces 
$\mathcal{F}$ and since such integrals can be interpreted as intersection
numbers between the cycles on $\mathcal{F}$, one can formally rewrite the
precursor to the Veneziano amplitude (as discussed in Part I) as 
\begin{equation}
I=\left( \frac{-\partial }{\partial f_{1}}\right) ^{r_{1}}\cdot \cdot \cdot
\left( \frac{-\partial }{\partial f_{d}}\right) ^{r_{d}}\int\limits_{\Delta
}\exp (\tilde{\Omega})\mid _{f_{i}=0\text{ }\forall i}=\int\limits_{\Delta }d%
\mathbf{x(}c_{1})^{r_{1}}\cdot \cdot \cdot (c_{d})^{r_{d}}  \tag{4.30}
\end{equation}%
provided that $r_{1}+\cdot \cdot \cdot +r_{d}=n$ . In such a language, the
problem of calculation of the Veneziano amplitudes using generating
function, Eq.(4.28), becomes mathematically almost equivalent to earlier
considered calculations related to the Witten-Kontsevich model discussed
earlier in Ref.[42].THis circumstance will be exploited in Part IV. Obtained
results provide complete symplectic solution of the Veneziano model.

As \ it was noticed by Atiyah and Bott [25], the replacement of exterior
derivative $d$ by $\tilde{d}=d+i(\xi )$ was inspired by earlier work by
Witten on supersymmetric formulation of quantum mechanics and Morse theory,
Ref.[26]. Such an observation allows us to discuss calculation of $\chi $
and, hence, the Veneziano amplitudes using the supersymmetric formalism
developed by Witten. The traditional way of developing Witten's ideas is
discussed in detail in earlier mentioned monograph, Ref.[70]. Its essence is
well summarized by Guillemin, Ref. [38]. Following this reference we notice
that $\chi $ is equal to the dimension $Q=Q^{+}-Q^{-}$ of the \ quantum
Hilbert space associated with the classical system described by the ( moment
map) Hamiltonian as discussed earlier in this section . To describe quantum
spaces associated with $Q^{+}$ and $Q^{-}$ \ we need to remind our readers
of several facts from the differential analysis on complex manifolds
discussed already in Part II..

We begin with the following observations. Let $X$ be the complex Hermitian
manifold and let $\mathcal{E}^{p+q}(X)$ denote the complex -valued
differential forms (sections) of type $(p,q)$ $,p+q=r,$ living on $X$. The
Hodge decomposition insures that $\mathcal{E}^{r}(X)$=$\sum\nolimits_{p+q=r}%
\mathcal{E}^{p+q}(X).$ The Dolbeault operators $\partial $ and $\bar{\partial%
}$ act on $\mathcal{E}^{p+q}(X)$ according to the rule \ $\partial :\mathcal{%
E}^{p+q}(X)\rightarrow \mathcal{E}^{p+1,q}(X)$ and $\bar{\partial}:\mathcal{E%
}^{p+q}(X)\rightarrow \mathcal{E}^{p,q+1}(X)$ , so that the exterior
derivative operator is defined as $d=\partial +\bar{\partial}$. Let now $%
\varphi _{p}$,$\psi _{p}\in \mathcal{E}^{p}$. By analogy with traditional
quantum mechanics we define (using Dirac's notations) the inner product 
\begin{equation}
<\varphi _{p}\mid \psi _{p}>=\int\limits_{M}\varphi _{p}\wedge \ast \bar{\psi%
}_{p}  \tag{4.31}
\end{equation}%
where the bar means the complex conjugation and the star $\ast $ means the
usual Hodge conjugation. Use of such a product is motivated by the fact that
the period integrals, e.g. those for the Veneziano-like amplitudes, and,
hence, those given by Eq.(4.28), are expressible through such inner products
[55]. Fortunately, such a product possesses properties typical for the
finite dimensional quantum mechanical Hilbert spaces. In particular, 
\begin{equation}
<\varphi _{p}\mid \psi _{q}>=C\delta _{p,q}\text{ and }<\varphi _{p}\mid
\varphi _{p}>>0,  \tag{4.32}
\end{equation}%
where $C$ is some positive constant. With respect to such defined scalar
product it is possible to define all conjugate operators, e.g. $d^{\ast }$,
etc. and, most importantly, the Laplacians 
\begin{align}
\Delta & =dd^{\ast }+d^{\ast }d,  \notag \\
\square & =\partial \partial ^{\ast }+\partial ^{\ast }\partial ,  \tag{4.33}
\\
\bar{\square}& =\bar{\partial}\bar{\partial}^{\ast }+\bar{\partial}^{\ast }%
\bar{\partial}.  \notag
\end{align}%
All this was known to mathematicians \textit{before} Witten's work [26%
\textbf{].} The unexpected twist occurred when Witten suggested to extend
the notion of the exterior derivative $d$. Within the de Rham picture (valid
for both real and complex manifolds) let $M$ be a compact Riemannian
manifold and $K$ be the Killing vector field which is just one of the
generators of isometry of $M,$ then Witten suggested to replace the exterior
derivative operator $d$ by the extended operator 
\begin{equation}
d_{s}=d+si(K)  \tag{4.34}
\end{equation}%
discussed earlier in the context of the equivariant cohomology. Here $s$ is
real nonzero parameter conveniently chosen. Witten argues that one can
construct the Laplacian (the Hamiltonian in his formulation) $\Delta $ by
replacing $\Delta $ by $\Delta _{s}=d_{s}d_{s}^{\ast }+d_{s}^{\ast }d_{s}$ .
This is possible if and only if $d_{s}^{2}=d_{s}^{\ast 2}$ $=0$ or, since $%
d_{s}^{2}=s\mathcal{L}(K)$ , where $\mathcal{L}(K)$ is the Lie derivative
along the field $K$, if the Lie derivative acting on the corresponding
differential form vanishes. The details are beautifully explained in the
much earlier paper by Frankel [68] mentioned earlier in Section 3.3. Atiyah
and Bott\ observed that the auxiliary multicomponent parameter \textbf{s}
can be identified with earlier introduced \textbf{f}. This observation
provides the link between the symplectic D-H formalism discussed earlier and
Witten's supersymmetric quantum mechanics. Looking at Eq.s (4.31) and
following Ref.s[3,38,39,70] we consider the (Dirac) operator $\acute{\partial%
}=\bar{\partial}+\bar{\partial}^{\ast }$ and its adjoint with respect to
scalar product, Eq.(4.30), then use of the above references suggests that 
\begin{equation}
Q=\ker \acute{\partial}-co\ker \acute{\partial}^{\ast }=Q^{+}-Q^{-}=\chi . 
\tag{4.35}
\end{equation}%
in accord with Vergne[3]. The results just described provide yet another
link between the supersymmetric and symplectic formalisms. Additional
details can be found both in Part II and references just cited.

\ 

\bigskip 

\textbf{Note added in proof}. After this work has been completed and
accepted for publication we become aware of the following two recent papers
: arxiv:mathCo/0507163 and arxiv: mathCO/0504231.These papers are not only
supporting results presented in the main text, they also provide numerous
additional details potentially useful for physical applications. Some of
these applications will be discussed in Part IV.

\bigskip\pagebreak

\bigskip

\textbf{References}

\bigskip

[1] \ \ \ A.Kholodenko, New string for old Veneziano amplitudes I.

\ \ \ \ \ \ \ Analytical treatment, J.Geom.Phys.55 (2005) 50-74.

[2] \ \ \ A.Kholodenko, New string for old Veneziano amplitudes II.

\ \ \ \ \ \ \ Group-theoretic treatment, J.Geom.Phys.(to be published).

[3] \ \ \ M. Vergne, Convex polytopes and quanization of symplectic
manifolds,

\ \ \ \ \ \ \ Proc.Natl.Acad.Sci. 93 (1996) 14238-14242.

[4] \ \ \ W. Lerche, C.Wafa, N.Warner, Chiral rings in N=2 superconformal

\ \ \ \ \ \ \ theories, Nucl.Phys. B324 (1989) 427-474.

[5] \ \ \ R. Stanley, Combinatorial reciprocity theorems,

\ \ \ \ \ \ \ Adv. Math. 14 (1974) 194-253.

[6] \ \ \ M. Brion, Points entiers dans les polyedres convexes,

\ \ \ \ \ \ \ Ann.Sci.Ecole Norm. Sup. 21 (1988) 653-663.

[7] \ \ A.Barvinok, A Course in Convexity, AMS Publishers,

\ \ \ \ \ \ \ Providence, RI, 2002.

[8] \ \ \ R.Diaz, S.Robins, The Ehrhart polynomial of a lattice polytope,

\ \ \ \ \ \ \ Ann.Math. 145 (1997) 503-518.

[9] \ \ \ R.Stanley, Combinatorics and Commutative Algebra,

\ \ \ \ \ \ \ Birkh\"{a}user, Boston, MA, 1996.

[10] \ V.Buchchtaber, T.Panov, Torus Actions and Their Applications

\ \ \ \ \ \ \ in Topology and Combinatorics, AMS Publishers,

\ \ \ \ \ \ \ Providence, RI, 2002.

[11] \ \ M.Green, J. Schwarz, E.Witten, Superstring Theory. Vol.1.,

\ \ \ \ \ \ \ Cambridge University Press, Cambridge,UK,1987.

[12] \ \ V. De Alfaro, S. Fubini, G.Furlan, C. Rossetti,

\ \ \ \ \ \ \ Currents in Hadron Physics, Elsevier Publ.Co., Amsterdam, 1973.

[13] \ R.Stanley, Invariants if finite groups and their applications to

\ \ \ \ \ \ \ \ combinatorics, BAMS (New Series) 1 (1979) 475-511.

[14] \ \ V.Batyrev, Variations of the mixed Hodge structure of affine

\ \ \ \ \ \ \ \ hypersurfaces in algebraic tori, Duke Math. Journal 69
(1993) 349-409.

[15] \ T.Hibi, Dual polytopes of rational convex polytopes,

\ \ \ \ \ \ \ Combinatorica 12 (1992) 237-240.

[16] \ N.Ashcroft, D.Mermin, Solid State Physics, Saunders Colledge Press,

\ \ \ \ \ \ Philadelphia, 1976.

[17] \ B.Greene, M.Plesser, Duality in Calabi-Yau moduli space,

\ \ \ \ \ \ \ Nucl.Phys. B338 (1990) 15-37.

[18] \ V.Batyrev, Dual polyhedra and mirror symmetry for Calabi-Yau

\ \ \ \ \ \ \ hypersurfaces in toric varieties, J.Alg.Geom. 3 (1994) 493-535.

[19] \ G.Veneziano, Construction of crossing-symmetric,Regge behaved,

\ \ \ \ \ \ \ amplitude for linearly rising tragectories,

\ \ \ \ \ \ \ Il Nuovo Chim. 57A (1968) 190-197.

[20] \ S.Donnachie, G.Dosch, P.Landshoff, O.Nachtmann, Pomeron Physics

\ \ \ \ \ \ and QCD, Cambridge U.Press, Cambridge, 2002.

[21] \ P.Collins, An Introduction to Regge Theory and High Energy Physics,

\ \ \ \ \ \ \ Cambridge U.Press, Cambridge, 1977.

[22] \ P.Frampton, Dual Resonance Models, W.A.Benjamin, Inc.,

\ \ \ \ \ \ \ Reading, Ma. , 1974.

[23] \ \ S.Mandelstam, Veneziano formula with trajectories spaced

\ \ \ \ \ \ \ by two units, Phys.Rev.Lett. 21 (1968) 1724-1728.

[24] \ A.Kholodenko, \ New models for Veneziano amplitudes: combinatorial,

\ \ \ \ \ \ \ symplectic and supersymmetric aspects,

\ \ \ \ \ \ \ Int.J.Geom.Methods in Mod.Physics 2 (2005) 563-584.

[25] \ M.Atiyah, R. Bott, The moment map and equivariant

\ \ \ \ \ \ \ cohomology, Topology 23 (1984)1-28.

[26] \ E. Witten, Supersymmetry and Morse theory,

\ \ \ \ \ \ \ J.Diff.Geom.17 (1982) 661-692.

[27] \ H. Coxeter, Regular Polytopes, The Macmillan Co., New York, 1963.

\bigskip \lbrack 28] \ G. Ziegler, Lectures on Polytopes, Springer-Verlag,
Inc., Berlin, 1995

[29] \ \ D.Ruelle, Dynamical Zeta Functions for Piecevise Monotone

\ \ \ \ \ \ \ \ Maps of the Interval\textit{, }AMS, Providence, RI, 1994.

[30] \ \ M. Atiyah, R. Bott, A Lefschetz fixed point formula for

\ \ \ \ \ \ \ \ elliptic complexes : I . Ann.Math.86 (1967) 374-407 ; ibid

\ \ \ \ \ \ \ \ A Lefschetz fixed point formula for elliptic complexes :

\ \ \ \ \ \ \ \ II.Applications. Ann.Math. 88 (1968) 451-491.

[31] \ \ I. Gelfand, M. Kapranov, A. Zelevinsky, Discriminants, Resultants

\ \ \ \ \ \ \ \ and Multidimensional Determinats,

\ \ \ \ \ \ \ \ Birkh\"{a}user, Inc., Boston, MA, 1994.

[32] \ \ N. Bourbaki, Groupes et Algebres de Lie (Chapitre 4-6),

\ \ \ \ \ \ \ Hermann, Paris, 1968.

[33] \ \ V.Guillemin, E. Lerman, S. Sternberg, Symplectic Fibrations

\ \ \ \ \ \ \ and Multiplicity Diagrams, Cambridge University Press,

\ \ \ \ \ \ \ Cambridge, UK,1996.

[34] \ \ P.Cartier, On Weil's character formula, BAMS 67 (1961) 228-230.

[35] \ \ R. Bott, On induced representations,

\ \ \ \ \ \ \ \ Proc.Symp.Pure Math.48 (1988) 1-13.

[36] \ \ V.Kac, Infinite Dimensional Lie Algebras,

\ \ \ \ \ \ \ Cambridge University Press, Cambridge, UK, 1990.

[37] \ R.Feres, Dynamical Systemsand Semisimple Groups: An Introduction,

\ \ \ \ \ \ \ Cambridge University Press, Cambridge, UK, 1998.

[38] \ V.Guillemin, Moment Maps and Combinatorial Invariants

\ \ \ \ \ \ \ of Hamiltonian T$^{n}$ Spaces, Birkh\"{a}user, Inc., Boston,
1994.

[39] \ V. Guillemin, V.Ginzburg, Y. Karshon, Moment Maps,

\textit{\ \ \ \ \ \ \ }Cobordisms and Hamiltonian Group Actions\textit{, }

\ \ \ \ \ \ \ AMS, Providence, RI, 2002.

[40]\ \ W.Fulton, Introduction to Toric Varieties, Ann.Math.Studies 131,

\ \ \ \ \ \ \ Princeton University Press, Princeton, 1993.

[41] \ G. Ewald, Combinatorial convexity and Algebraic Geometry,

\ \ \ \ \ \ \ Springer-Verlag, Inc., Berlin, 1996.

[42] \ V. Danilov, The geometry of toric varieties, Russ.Math.Surveys 33

\ \ \ \ \ \ \ (1978)\ \ 97-154.

[43] \ \ B.Iversen, The geometry of algebraic groups,

\ \ \ \ \ \ \ Adv.Math.20 (1976) 57-85.

[44] \ \ B.Iversen, H.Nielsen, Chern numbers and diagonalizable groups,

\ \ \ \ \ \ \ \ J.London Math.Soc. 11 (1975) 223-232.

[45] \ \ H.Hopf, Uber die topologie der gruppen-manigfaltigkeiten uber ihre

\ \ \ \ \ \ \ \ verallgemeinerungen, Ann.Math. 42 (1941) 22-52.

[46] \ \ H.Hopf, H.Samelson, Ein satz uber die wirkungesraume geschlossener

\ \ \ \ \ \ \ \ Liescher gruppen, Comm.Math.Helv. 13 (1941) 240-251.

[47] \ A.Borel, Linear Algebraic Groups\textit{,} Springer-Verlag, Inc.,
Berlin, 1991.

[48] \ I. Macdonald, Linear Algebraic Groups, LMS Student Texts 32,

\ \ \ \ \ \ \ Cambridge University Press, Cambridge, UK,1999.

[49] \ J. Humphreys, Linear Algebraic Groups\textit{,}

\ \ \ \ \ \ \ \ Springer-Verlag, Inc., Berlin, 1975.

[50] \ H. Hiller, Geometry of Coxeter Groups, Pitman Inc., Boston, 1982.

[51] \ A.Kholodenko, Kontsevich-Witten model from 2+1 gravity:

\ \ \ \ \ \ \ new exact combinatorial solution. J.Geom.Phys. 43 (2002) 45-91.

[52] \ F. Knudsen, G. Kempf, D.Mumford, B.Saint-Donat,

\ \ \ \textit{\ \ \ \ }Toroidal Embeddings I, LNM\ 339\textbf{,}

\ \ \ \ \ \ \ Springer-Verlag, Inc., Berlin, 1973.

[53] \ J-P. Serre, Algebres de Lie Semi-Simples Complexes,

\ \ \ \ \ \ \ Benjamin, Inc., New York, 1966.

[54] \ J.Humphreys, Introduction to Lie Algebras and Representation\textit{\ 
}

\ \ \ \ \ \ \ Theory\textit{,} Springer-Verlag, Inc., Berlin, 1972.

[55] \ R.Wells, Differential Analysis on Complex Manifolds,

\ \ \ \ \ \ \ Springer-Verlag, Inc., Berlin, 1980.

[56] \ V.Ginzburg, Representation Theory and Complex Geometry\textit{,}

\ \ \ \ \ \ \ Birkh\"{a}user Verlag, Inc., Boston, 1997.

[57] \ A. Kirillov, Elements of the Theory of Representations\textit{,} (in
Russian),

\ \ \ \ \ \ \ Nauka, Moscow, 1972.

[58] \ A. Fomenko, V. Trofimov, Integrable Systems on Lie Algebras

\ \ \ \textit{\ \ \ \ }and Symmetric Spaces,

\ \ \ \ \ \ \ Gordon and Breach Publishers, New York,1988.

[59] \ A. Kholodenko, Use of meanders and train tracks for description of

\ \ \ \ \ \ \ defects and textures in liquid crystals and 2+1 gravity,
J.Geom.Phys.

\ \ \ \ \ \ \ 33 (2000) 23-58.

[60] \ M. Atiyah, Angular momentum, convex polyhedra and algebraic

\ \ \ \ \ \ \ \ geometry, Proceedings of the Edinburg Math.Society 26 (1983)
121-138.

[61] \ M. Audin, Torus Actions on Symplectic Manifolds\textit{,}

\ \ \ \ \ \ \ Birkh\"{a}user, Inc., Boston, 2004.

[62] \ M.Atiyah, Convexity and commuting Hamiltonians,

\ \ \ \ \ \ \ Bull.London Math.Soc.14 (1982) 1-15.

[63] \ A. Schrijver, Combinatorial Optimization. Polyhedra and Efficiency%
\textit{.}

\ \ \ \ \ \ \ \ Springer-Verlag, Inc., Berlin, 2003.

[64] \ S. Gass, Linear Programming, McGraw Hill Co., New York, 1975.

[65] \ O.Debarre, Fano Varieties. In Higher Dimensional Varieties and

\ \ \ \ \ \ \ Rational Points, pp.93-132, Springer-Verlag, Berlin, 2003.

[66] \ V. Guillemin, S. Sternberg, Convexity properties of the moment

\ \ \ \ \ \ \ mapping, Invent.Math. 67 (1982) 491-513.

[67] \ T. Delzant, Hamiltoniens periodiques et image convexe de

\ \ \ \ \ \ \ l'application moment, Bull.Soc.Math.France 116 (1988) 315-339.

[68] \ T. Frankel, Fixed points and torsion on K\"{a}hler manifolds.

\ \ \ \ \ \ \ Ann.Math. 70 (1959) 1-8.

[69] \ H. Flaska, Integrable systems and torus actions,

\ \ \ \ \ \ \ In O.Babelon, P.Cartier, Y.Schwarzbach (Eds)

\ \ \ \ \ \ \ Lectures on Integrable Systems,

\ \ \ \ \ \ \ World Scientific Pub.Co., Singapore, 1994.

[70] \ N.Berline, E.Getzler, M.Vergne, Heat Kernels and Dirac Oprators,

\ \ \ \ \ \ \ Springer-Verlag, Berlin, 1991.

[71] \ A. Kholodenko, New Veneziano amplitudes from \textquotedblright
old\textquotedblright\ Fermat

\ \ \ \ \ \ \ (hyper) surfaces. In C.Benton (Ed):

\ \ \ \ \ \ \ Trends in Mathematical Physics

\textit{\ \ \ \ \ \ }Research\textit{,} pp\textit{\ 1-94, }Nova Science
Publ., New York, 2004.

[72] \ M.Vergne, Residue formulae for Verlinde sums, and for number

\ \ \ \ \ \ \ of integral points in convex polytopes. In E.Mezetti, S.Paycha
(Eds),

\ \ \ \ \ \ \ European Women in Mathematics, pp 225-284,

\ \ \ \ \ \ \ World Scientific, Singapore, 2003.

[73] M.Brion, M.Vergne, Lattice points in simple polytopes,

\ \ \ \ \ \ J.AMS 10 (1997) 371-392.

[74] \ A. Khovanskii, A.Pukhlikov, A Riemann-Roch theorem for integrals

\ \ \ \ \ \ \ and sums of quasipolynomials over virtual polytopes,

\ \ \ \ \ \ \ St.Petersburg Math.J. 4 (1992) 789-812.

[75] \ A.Barvinok, K.Woods, Short rational generating functions for

\ \ \ \ \ \ \ lattice point problems, AMS Journal, 16 (2003) 957-979.

[76] \ M.Brion, M.Vergne, An Equivariant Riemann-Roch theorem

\ \ \ \ \ \ \ for complete, simplicial toric varieties,

\ \ \ \ \ \ \ J.Reine Angew. Math.482 (1997) 67-92.

[77] \ F.Hirzebruch, D.Zagier, The Atiyah-Singer Theorem and Elementary

\ \ \ \ \ \ \ Number Theory, Publish or Perish Inc., Berkeley, Ca, 1974.

[78] \ V.Guillemin, S.Sternberg, Supersymmetry and Equivariant de\ Rham

\ \ \ \ \ \ \ Theory, Springer-Verlag, Berlin, 1999.

\bigskip

\ \ \ 

\bigskip

\bigskip

\ \ \ \ \ \ \ \ \ \ \ \ 

\ 

\ \ \ \ \ \ \ \ \ \ \ 

\bigskip

\bigskip

\bigskip

\end{document}